\documentclass{emulateapj}
\usepackage{tablefootnote}
\usepackage{amsmath}

\begin{document}

\title{Hydrodynamic escape of water vapor atmospheres near very active stars}

\author{C. P. Johnstone \altaffilmark{1}}
\affil{University of Vienna, Department of Astrophysics, T\"{u}rkenschanzstrasse 17, 1180 Vienna, Austria}
\email{colin.johnstone@univie.ac.at}

\begin{abstract}
When exposed to the high energy X-ray and ultraviolet radiation of a very active star, water vapor in the upper atmospheres of planets can be photodissociated and rapidly lost to space. 
In this paper, I study the chemical, thermal, and hydrodynamic processes in the upper atmospheres of terrestrial planets, concentrating on water vapor dominated atmospheres orbiting in the habitable zones of active stars.
I consider different stellar activity levels and find very high levels of atmospheric escape in all cases, with the outflowing gas being dominated by atomic hydrogen and oxygen in both their neutral and ion forms.
In the lower activity cases, I find that the accumulation of O$_2$ and increases in the D/H ratios in the atmospheres due to mass fractionation are possible, but in the higher activity cases no mass fractionation takes place.
Connecting these results to stellar activity evolution tracks for solar mass stars, I show that huge amounts of water vapor can be lost, and both the losses and the amount of O$_2$ that can be accumulated in the atmosphere depend sensitively on the star's initial rotation rate.
For an Earth-mass planet in the habitable zone of a low-mass M-dwarf, my results suggest that the accumulation of atmospheric O$_2$ is unlikely unless water loss can take place after the star's most active phase.
\end{abstract}


\section{Introduction}

\footnotetext{
Tabulated output data from all atmosphere simulations presented in this paper, accompanied by Python scripts used for making all figures, can be downloaded from \url{https://zenodo.org/record/3576049\#.XfTa-tl7lhE}.  
}

Water is one of the most important molecules for planets and can influence significantly their surfaces, climates, and potential for sustaining life. 
In addition to being present in the atmospheres of solar system planets, water has been detected in the atmospheres of many exoplanets (e.g. \citealt{Tinetti07}; \citealt{Wakeford13}; \citealt{Brogi14}; \citealt{Tsiaras19}).
A large number of physical processes take place in the upper atmospheres and exospheres of planets that cause water to be lost to space, most of which are driven by interactions with the central star. 
For example, non-thermal escape processes include ion pick-up by stellar winds (\citealt{Kislyakova14a}), cold ion outflows (\citealt{Glocer07}), and the loss of photochemically produced high-energy particles (\citealt{Amerstorfer17}).
Thermal escape mechanisms include hydrodynamic escape, which takes place when the upper atmosphere is heated to such a temperature that the thermal pressure causes it to accelerate away from the planet at speeds exceeding the escape velocity.
This can happen without any outside influences for very low-mass planets (\citealt{Stoekl15}), in response to the star's bolometric radiation for planets on short period orbits (\citealt{OwenWu16}), or most notably in response to the star's X-ray and ultraviolet emission for planets orbiting active stars (\citealt{Tian05}; \citealt{Erkaev16}; \citealt{Kubyshkina18}).
Heating can also take place due to other mechanisms, mostly in response to the star's wind (\citealt{Cohen14}; \citealt{Lichtenegger16}).
For example, \citet{Chassefiere96} and \citet{Chassefiere97} suggested for the case of early Venus that energetic neutral atoms (ENAs) created in charge exchange reactions between neutral atmospheric particles and the early solar wind could heat the atmosphere significantly and drive escape, though \citet{Lichtenegger16} found that this mechanism might not lead to significant additional escape.

Water vapor in a planet's upper atmosphere is photodissociated by the high-energy radiative spectrum of its host star creates a large number of other chemical species such as OH, O$_2$, and O$_3$, and causes most of the thermosphere to be filled neutral and ionized hydrogen and oxygen atoms which can flow away from the planet hydrodynamically (\citealt{Kasting83}; \citealt{Guo19}).
Due to its lower mass, hydrogen is lost more rapidly than oxygen, though how different these loss rates are depends on several factors, including the total loss rate and which loss process dominates.
This preferential loss of H has been suggested as a mechanism for producing significant amounts of O$_2$ in an atmosphere (\citealt{WordsworthPierrehumbert14}; \citealt{LugerBarnes15}).
In this scenario, a planet with a large amount of atmospheric H$_2$O orbiting a very active star will undergo rapid escape but with much of the O remaining, leading to O$_2$ build-up.  
For potentially habitable planets, it is expected that this build-up might be more significant for planets orbiting M dwarfs since these stars spend a very long time on the pre-main-sequence, meaning that planets that will eventually be in the habitable zones when the star reaches the main-sequence spend a long time inside the inner edge of the habitable zone where they are too hot for liquid surface water (\citealt{RamirezKaltenegger14}).

The build-up of O$_2$ can be prevented if oxygen is absorbed into the surface, which \citet{Wordsworth18} showed could significantly reduce the build-up of O$_2$ during a planet's early magma ocean phase. 
In this case the absorbed oxygen can still be released into the atmosphere later in the planet's lifetime.
Alternatively, as the O$_2$ accumulates, the mixing ratio of O in the upper atmosphere will increase, leading to a decrease in the difference between the H and O loss rates (\citealt{Tian15}).
The loss of water can also be inhibited by the presence of a cold-trap if non-condensing species such as N$_2$ or CO$_2$ make up a large fraction of the atmosphere (\citealt{WordsworthPierrehumbert13}).
An initially pure H$_2$O atmosphere cannot have a cold-trap, but a cold-trap can form as O$_2$ builds up, inhibiting H$_2$O escape (\citealt{WordsworthPierrehumbert14}).
If the star's activity is strong enough to drive rapid atmospheric escape however, we might expect rapid oxygen escape to take place in such a case, possibly removing any accumulated atmospheric O$_2$.  
Studying the case of GJ~1132b, which is an Earth-mass planet orbiting a low mass M dwarf, \citet{Schaefer16} found that most of the O$_2$ produced by the dissociation and loss of water vapor is lost to space, whereas a much smaller fraction is absorbed into the planet.

Most atmospheric loss processes take place in response to the magnetic activity of the host star, which is responsible for winds and high-energy X-ray and ultraviolet radiation.
A star's X-ray and ultraviolet spectrum (referred to here as `XUV' and defined as 1 to 400~nm) is emitted by the photosphere and the magnetically heated chromosphere and corona (\citealt{Fontenla16}).
For a solar mass star, the chromosphere and corona is responsible for the emission at wavelengths shorter than $\sim$200~nm and this radiation is absorbed high in the atmospheres of planets, driving heating and photochemistry.
This emission depends sensitively on the rotation rate of the star (\citealt{Reiners14}).
Since rotation rates decline with age due to angular momentum removal by stellar winds, there is also a corresponding decline in XUV emission (\citealt{Guedel97}).
This means that when the solar system was young the upper atmospheres of solar system planets were likely significantly hotter and more expanded than currently (\citealt{Kulikov07}; \citealt{Tian08}; \citealt{Johnstone18}).
However, the situation is complicated by the fact that stars can be born with different rotation rates, and stars born as slow rotators will evolve very differently to stars born as fast rotators (\citealt{Johnstone15}; \citealt{Tu15}).
A solar mass star born as a rapid rotator will remain highly active for much longer than a star born as a slow rotator, which has important consequences for the subsequent evolution of planetary atmospheric escape (\citealt{Johnstone15letter}).

In this paper, I study the reaction of a water vapor atmosphere to the high XUV spectrum of a very active star.
The aim is to use a newly developed state-of-the-art physical upper atmosphere model to study hydrodynamically outflowing water vapor atmospheres and to understand the loss rates of hydrogen, deuterium, and oxygen.
In Section~\ref{sect:model}, I describe the physical upper atmosphere model used here, in Section~\ref{sect:results}, I present the modelling results, in Section~\ref{sect:evo}, I use these results to study the long term evolution of water losses, and in Section~\ref{sect:conclusions}, I discuss the significance of the results.

\begin{figure}
\centering
\includegraphics[width=0.45\textwidth]{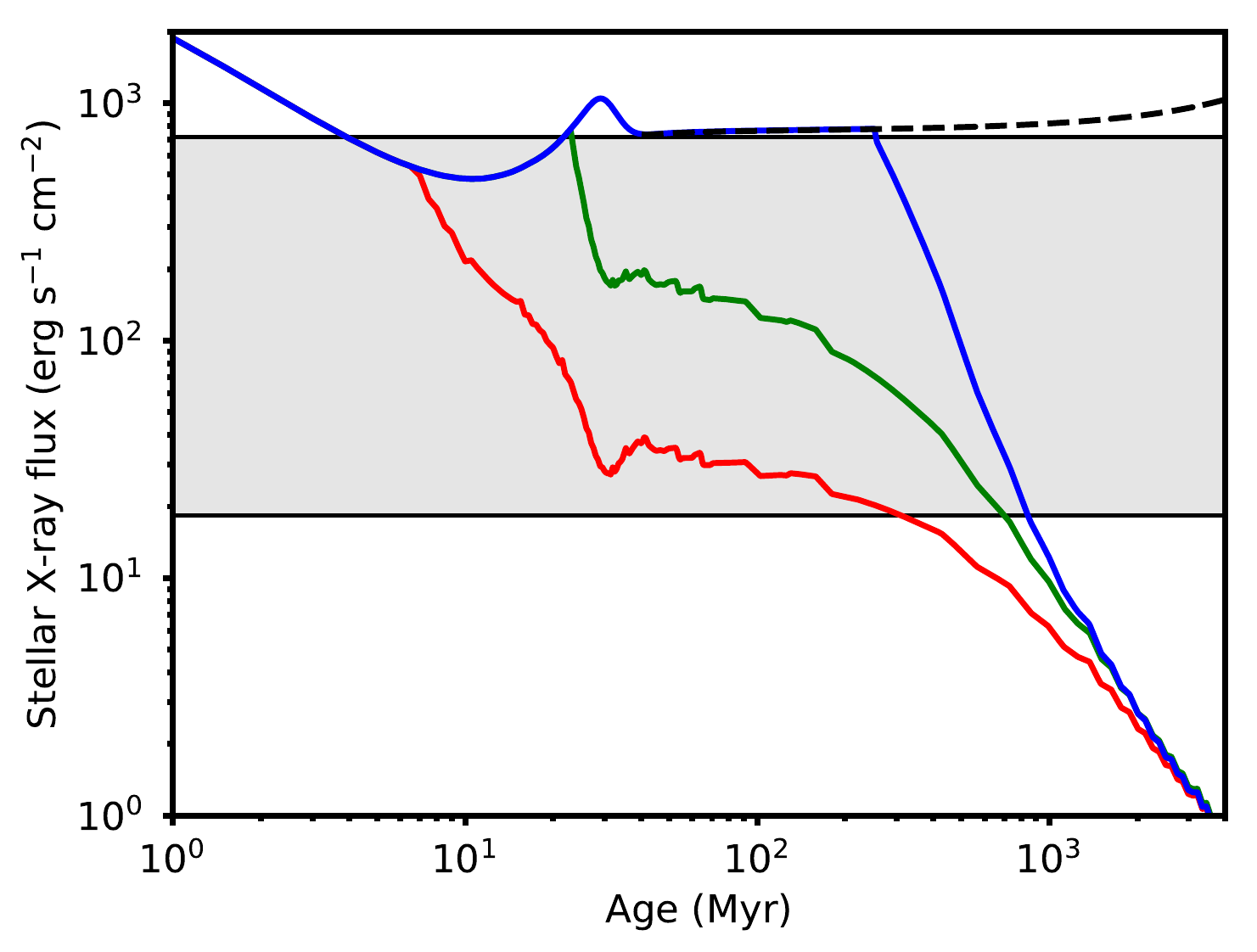}
\caption{
Evolution of stellar X-ray flux at 1~AU for solar mass stars with different initial rotation rates as calculated by \citet{Tu15}.
The red, green, and blue tracks correspond to the cases of slow, median, and fast rotators and the dashed black line shows the case of a star that remains at the saturation threshold for its entire lifetime.
The grey shades area shows the range of activity levels that I study in this paper.
}
\label{fig:XUVtracks}
\end{figure}

\section{Model} \label{sect:model}

\begin{figure}
\centering
\includegraphics[width=0.45\textwidth]{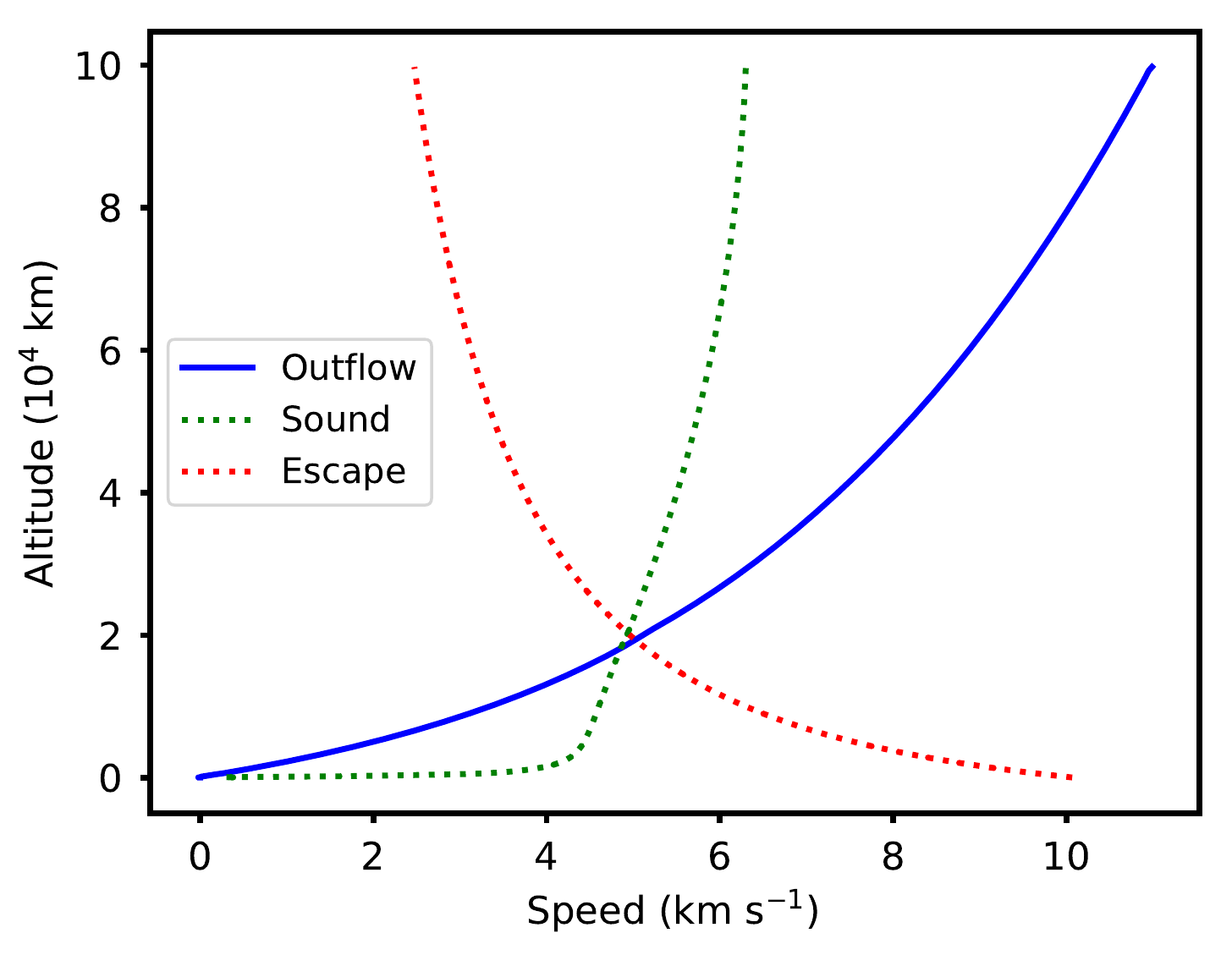}
\includegraphics[width=0.45\textwidth]{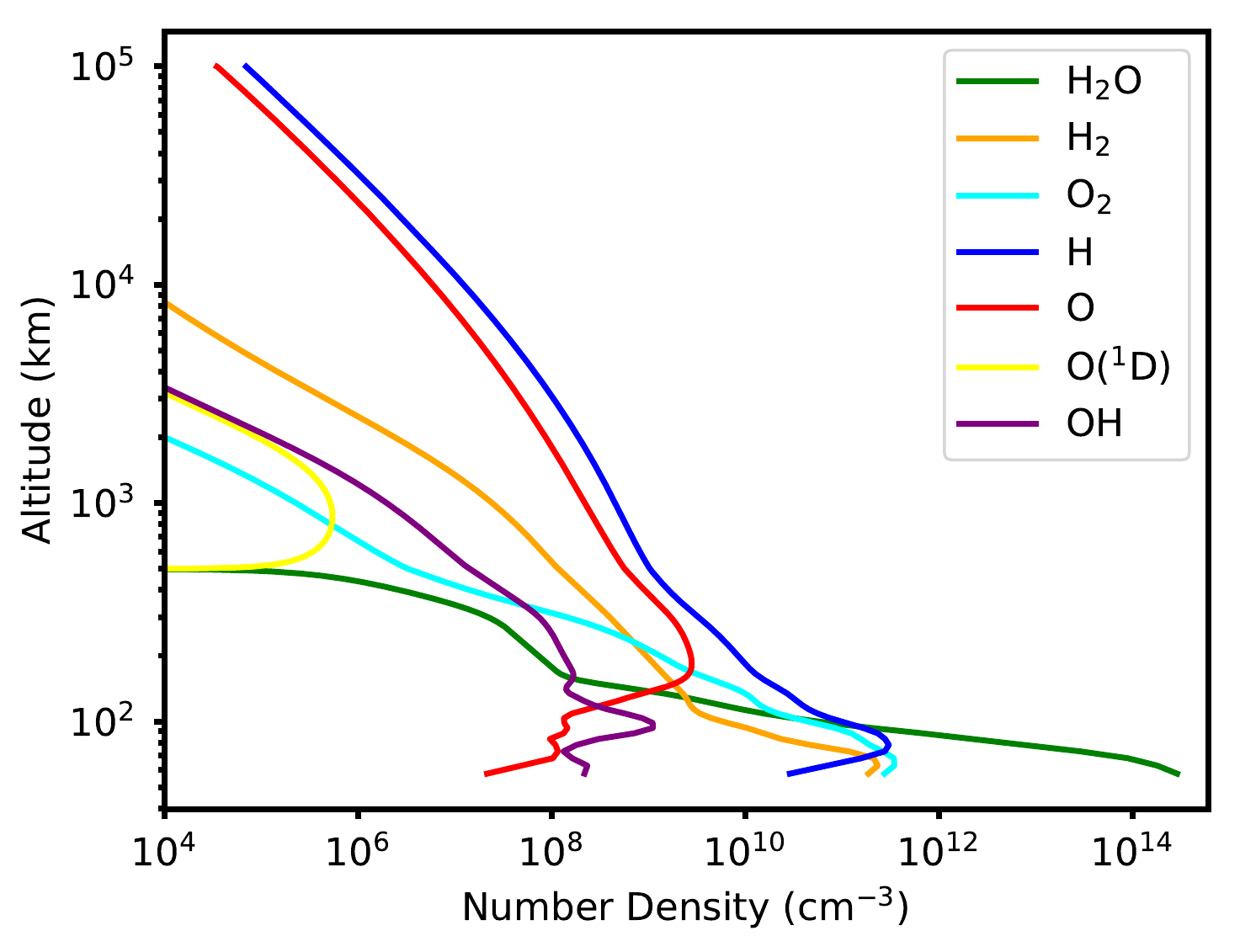}
\includegraphics[width=0.45\textwidth]{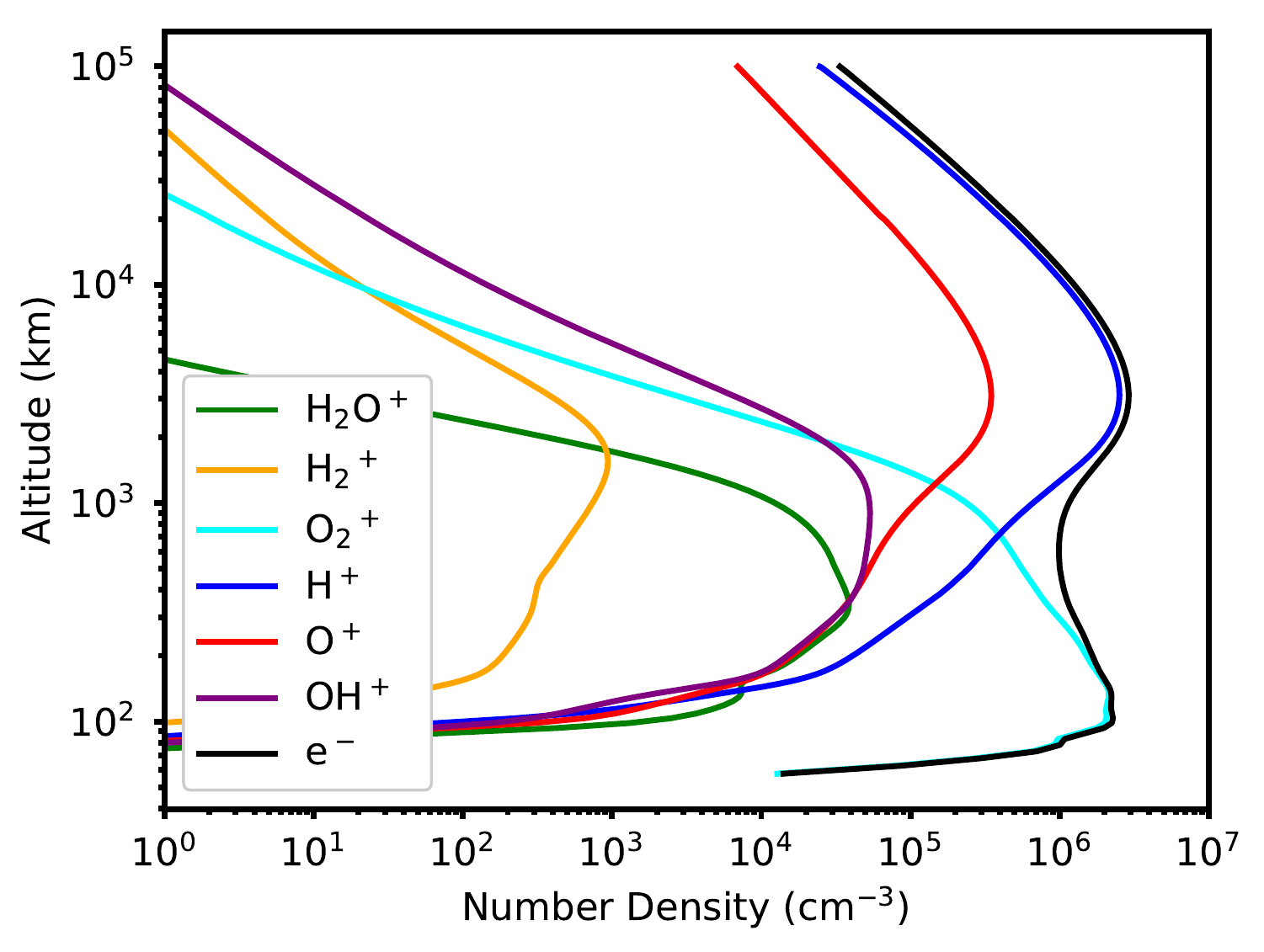}
\caption{
Hydrodynamic simulation of a rapidly outflowing water vapor atmosphere.
The panels show as functions of altitude the outflow speed (\emph{upper-panel}) and the densities of selected neutral (\emph{middle-panel}) and ion (\emph{lower-panel}) species. 
In the upper panel, the sound and escape speeds are shown and all three lines cross at the same location, as required for a pressure driven Parker wind.
}
\label{fig:Case5}
\end{figure}

\begin{deluxetable*}{ccccccccccccc}
\centering
\tabletypesize{\footnotesize}
\tablewidth{0.99\textwidth}
\tablecolumns{12}
\tablecaption{Basic properties of the cases studied here.}
\tablehead{
\colhead{Case} &
\colhead{$F_\mathrm{X}^1$} &
\colhead{$F_\mathrm{EUV}$} &
\colhead{$F_\mathrm{FUV}$} &
\colhead{$\Delta F_\mathrm{XUV}^2$} &
\colhead{Solar age$^3$} &
\colhead{$T_\mathrm{max}$} &
\colhead{$v_\mathrm{max}$} &
\colhead{$z_\mathrm{sonic}$} &
\colhead{$X_\mathrm{ion,max}$} &
\colhead{$10^{12} F_\mathrm{mass}$} &
\colhead{$f_\mathrm{H,D}$} &
\colhead{$f_\mathrm{H,O}$} \\
\colhead{} &
\multicolumn{4}{c}{(erg~s$^{-1}$~cm$^{-2}$)} &
\colhead{(Myr)} &
\colhead{(10$^3$~K)} &
\colhead{(km~s$^{-1}$)} &
\colhead{(10$^4$~km)} &
\colhead{} &
\colhead{(g~s$^{-1}$~cm$^{-2}$)} &
\colhead{} &
\colhead{} 
}
\startdata
1 & 18.4 & 98.7 & 231.3 & 256.2 & 310-840 		& 3.71 	& 6.21 & 2.25 & 0.075 & 0.272 & 0.861 & 0.402 \\
2 & 22.9 & 124.5 & 252.5 & 302.6 & 180-790		& 4.78 	& 6.87 & 2.00 & 0.091 & 0.367 & 0.853 & 0.494 \\
3 & 30.0 & 166.0 & 280.9 & 373.4 & 93-730 		& 6.38 	& 7.73 & 1.72 & 0.12 & 0.529 & 0.868 & 0.613 \\
4 & 42.3 & 241.2 & 321.5 & 494.2 & 26-640 		& 8.98 	& 8.96 & 1.43 & 0.16 & 0.829 & 0.903 & 0.765 \\
5 & 68.5 & 409.8 & 391.6 & 746.9 & 22-550 		& 13.3 	& 11.0 & 1.11 & 0.24 & 1.46 & 0.953 & 0.925 \\
6 & 77.5 & 470.2 & 412.8 & 833.9 & 22-530 		& 14.1 	& 11.5 & 1.04 & 0.27 & 1.66 & 0.962 & 0.951 \\
7 & 89.0 & 548.2 & 438.0 & 944.2 & 20-510 		& 15.3 	& 12.1 & 0.964 & 0.29 & 1.92 & 0.971 & 0.972 \\
8 & 104.0 & 651.8 & 468.5 & 1088.0 & $<$490 	& 16.3 	& 12.8 & 0.875 & 0.32 & 2.45 & 0.980 & 0.990 \\
9 & 124.2 & 794.7 & 506.6 & 1282.6 & $<$460 	& 17.4 	& 13.5 & 0.794 & 0.37 & 2.67 & 0.986 & 1.000 \\
10 & 152.8 & 1001.7 & 555.5 & 1558.6 & $<$440 	& 18.3 	& 14.3 & 0.729 & 0.43 & 3.25 & 0.993 & 1.000 \\
11 & 195.7 & 1321.9 & 621.4 & 1976.0 & $<$400 	& 19.3 	& 15.3 & 0.650 & 0.48 & 3.76 & 0.994 & 1.000 \\
12 & 266.3 & 1866.0 & 715.8 & 2667.5 & $<$360 	& 19.5 	& 16.3 & 0.560 & 0.59 & 4.95 & 0.998 & 1.000 \\
13 & 399.4 & 2933.9 & 865.2 & 3987.3 & $<$320 	& 19.0 	& 17.5 & 0.524 & 0.75 & 7.03 & 1.000 & 1.000 \\
14 & 719.0 & 5632.0 & 1146.9 & 7217.5 & $<$250 	& 15.1 	& 19.1 & 0.482 & 0.93 & 11.3 & 1.000 & 1.000
\enddata
\tablenotetext{1}{
For the fluxes at 1~AU, I define the wavelength ranges for $F_\mathrm{X}$, $F_\mathrm{EUV}$, and $F_\mathrm{FUV}$ as $<$10~nm, 10-100~nm, and 100-200~nm respectively.
}
\tablenotetext{2}{
The quantity $\Delta F_\mathrm{XUV}$ is the difference in the total XUV flux in the wavelength range considered in this paper (1-400~nm) between the upper and lower boundaries of the simulation domain. 
}
\tablenotetext{3}{
The solar ages give the range of ages when the Sun, or any Sun-like star, is likely to have this average activity level based on the 10th and 90th percentile X-ray evolutionary tracks calculated by \citet{Tu15}.
}
\label{table:grid}
\end{deluxetable*}

The system that I study in this paper consists of an Earth-mass planet with a pure water vapor atmosphere orbiting at 1~AU around an active solar mass star. 
Using my atmosphere model, I calculate the 1D atmosphere structure between altitudes of 50 and 100,000~km assuming a zenith angle of 0$^{\circ}$.
The zero zenith angle assumption is motivated by the assumptions made in Section~\ref{sect:evo} to calculate total atmospheric loss rates from the model results.
The lower boundary altitude is largely arbitrary and our results are not sensitive to this choice since it is anyway negligible compared to the radius of the planet. 
At the lower boundary, I assume a gas composed of H$_2$O and HDO, with a temperature of 250~K, corresponding approximately to the planet's effective temperature, and a number density of \mbox{$5 \times 10^{14}$~cm$^{-3}$}.
The mixing ratio of HDO at the lower boundary is \mbox{$1.6 \times 10^{-4}$}, corresponding approximately to the value in the Earth's oceans (\citealt{Eberhardt95}) and is typical for values in chondritic meteorites (\citealt{Marty16}), though since the abundance of deuterium is too small to influence the model, any sufficiently small base HDO mixing ratio could have been chosen.
The base number density is mostly arbitrary and experiments with other values have shown that my results are not sensitive to this value as long as it is large enough that all of the absorption of the XUV spectrum takes place within the computational domain.  
The upper boundary altitude is also arbitrary and only chosen to be far enough from the planet that the hydrodynamic outflow is supersonic within the simulation domain; at this boundary, standard zero-gradient outflow conditions are used. 
 
For the stellar XUV spectra in the wavelength range 1 to 400~nm, I use the method and codes developed for solar mass stars by \citet{Claire12}.
It is important to use realistic stellar spectra since the shape of a star's XUV spectrum depends on its activity level: more active stars have hotter coronae (\citealt{JohnstoneGuedel15}), meaning that larger fractions of their X-ray and extreme ultraviolet emission are at the higher energy parts of the spectrum (\citealt{Guedel04}; \citealt{SanzForcada11}).
In Fig.~\ref{fig:XUVtracks}, I show evolutionary tracks for the X-ray emission of solar mass stars with different initial rotation rates, where the shades region shows the range of activity levels that I study in this paper.
In this paper, I use the X-ray flux, $F_\mathrm{X}$, at the planet's orbit as the measure of stellar activity because the X-ray luminosity of a star is easily measured and often available for exoplanet hosts.

To calculate the atmospheric structures and loss rates, I use The Kompot Code, recently developed by \citet{Johnstone18} and \citet{Johnstone19a}. 
This is a sophisticated general-purpose first-principles physical model for the upper atmospheres of planets, developed to take into account the range of physical processes taking place in the upper atmosphere. 
The model has been designed such that it can be applied to any type of planet with arbitrary atmospheric compositions.
In \citet{Johnstone18}, detailed descriptions of the physical model (Section~2) and numerical methods (Appendices) can be found. 
As in \citet{Johnstone19a}, I have simplied slightly the model presented in \citet{Johnstone18} by removing the assumption that the neutrals, ions, and electrons have separate temperatures and instead only using a single temperature.
I have tested this simplification by rerunning the model for the modern Earth's upper atmosphere and find good agreement with the models presented in \citet{Johnstone18}.
Another difference to the model is the chemical network which I have changed in two ways.
Firstly, all chemical reactions and species that contain elements other than hydrogen and oxygen have been removed since I assume a purely H$_2$O gas at the lower boundary.
Secondly, I have added chemical reactions and species involving deuterium; the reactions included in the model are taken from \citet{LiangYung09} and \citet{GarciaMunoz07}, with additional XUV photoreactions taken from the PHIDRATES database (\citealt{HuebnerMukherjee15}). 

The model starts from a set of arbitrary initial conditions and evolves the state of the system forward in time by a large number of small timesteps until it reaches a steady state.
The basic set of equations describing the evolution of the atmospheric properties is
\begin{equation} \label{eqn:main_speciescontinuity}
\frac{\partial n_j}{ \partial t} 
+ \frac{1}{r^2} \frac{\partial \left[ r^2 ( n_j v + \Phi_{\mathrm{d},j}) \right] }{\partial r} 
=
S_j ,
\end{equation}
\begin{equation} \label{eqn:main_momentum}
\frac{\partial ( \rho v ) }{ \partial t} 
+ \frac{1}{r^2} \frac{\partial \left[ r^2 \left( \rho v^2 + p \right) \right] }{\partial r} 
=
- \rho g 
+ \frac{2 p }{r} ,
\end{equation}
\begin{equation} \label{eqn:main_energy}
\begin{aligned}
\frac{\partial e }{ \partial t} &
+ \frac{1}{r^2} \frac{\partial \left[ r^2 v \left( e + p \right) \right] }{\partial r} 
= - \rho v g + Q_{\mathrm{h}} - Q_{\mathrm{c}} \\
 & + \frac{1}{r^2} \frac{\partial}{\partial r} \left[ r^2 \kappa_\mathrm{cond} \frac{\partial T}{\partial r} + r^2 \kappa_\mathrm{eddy} \left( \frac{\partial T}{\partial r} + \frac{g}{c_\mathrm{P}} \right) \right] ,
\end{aligned} 
\end{equation}
where
$r$ is the radius,
$n_j$ is the number density of the $j$th species,
$\rho$ is the total mass density,
$v$ is the bulk advection speed,
\mbox{$\rho v$} is the momentum density,
$e$ is the energy density,
$T$ and $p$ are the temperature and thermal pressure,
$\Phi_{\mathrm{d},j}$ and $S_j$ are the diffusive particle flux and chemical source term of the $j$th species,
$g$ is the gravitational acceleration,
$Q_{\mathrm{h}}$ and $Q_{\mathrm{c}}$ are the volumetric heating and cooling rates,
$\kappa_\mathrm{cond}$ is the thermal conductivity,
$\kappa_\mathrm{eddy}$ is the eddy conductivity,
and $c_\mathrm{P}$ is the specific heat at constant pressure.
Since chemistry and diffusion do not change the total mass density of the gas, Eqn.~\ref{eqn:main_speciescontinuity} implies the standard mass continuity equation.

The evolution of the chemical structure of the atmosphere is described by Eqn.~\ref{eqn:main_speciescontinuity}.
The processes taken into account are hydrodynamic advection, eddy and molecular diffusion, and chemistry, including neutral and ion chemistry and XUV driven photochemistry.
In total, 223 chemical reactions, including 32 photoreactions, are considered, and the gas is composed of 33 chemical species of which 14 are ions. 
The chemistry is solved in a time-dependent way with no assumption of chemical equilibrium. 
The full set of hydrodynamic equations are solved using the explicit scheme described in \citet{Kaeppeli16}.
This scheme is `well-balanced', and can therefore accurately calculate the structure of an atmosphere that is in hydrostatic equilibrium.
The evolution of the thermal structure of the atmosphere is described by Eqn.~\ref{eqn:main_energy}.
The processes taken into account are hydrodynamic advection, heating by stellar XUV and infrared radiation, cooling by the emission of radiation to space, and thermal conduction.
The effects of adiabatic cooling are present in the model.
A strength of my model is the calculation of the heating rate, which is done from first-principles without the use of arbitrary free parameters such as the heating efficiency. 
The XUV heating model includes the effects of direct heating by absorbed XUV photons, energy released by exothermic chemical reactions, and heating of thermal electrons by collisions with high-energy photoelectrons created by photoionization reactions.

The cooling model includes the emission of radiation to space by H$_2$O, H, and O. 
For H$_2$O cooling, I use the method by \citet{Hollenbach79} and summarized by \citet{Kasting83} who used this method for calculating cooling from water vapor in planetary atmospheres; this model considers excitation of water molecules by atomic hydrogen and takes into account non-local thermodynamic equilibrium effects.
For H cooling, I consider the effects of Ly-$\alpha$ emission to space using the method given by \citet{MurrayClay09} and \citet{Guo19}.
For O cooling, I consider emission at 63~$\mu$m and 147~$\mu$m using the parameterisations derived by \citet{Bates51} and commonly used for calculating cooling in the upper thermospheres of planets.
In future work, I intend to implement in the model a more sophisticated treatment of radiative cooling.

Another change to the model that I make for this study is the way that the molecular diffusion coefficients are calculated. 
Molecular diffusion coefficients are different for different chemical species and depend on the composition of the background gas. 
In previous models, I used measured diffusion coefficients for several common species (see Eqn.~25 of \citealt{Johnstone18}) assuming a background atmosphere composed of N$_2$, and then made reasonable guesses for the values for all other species. 
As in \citet{Hobbs19}, I use in this paper the equation for the molecular diffusion coefficients for the $i$th species, $D_i$, derived by \cite{ChapmanCowling70}, given by
\begin{equation}
D_i = \frac{3}{8 N \left[ 0.5 ( d_i + \bar{d} ) \right]^2} \left[ \frac{k_\mathrm{B} T \left( m_i + \bar{m} \right) }{ 2 \pi m_i \bar{m} } \right]^{\frac{1}{2}}
\end{equation}
where $N$ is the total number density of the gas, $d_i$ is the particle diameter of the $i$th species\footnotemark, $\bar{d}$ is the average particle diameter of the entire gas, $k_\mathrm{B}$ is the Boltzmann constant, $T$ is the gas temperature, $m_i$ is the molecular mass of the $i$th species, and $\bar{m}$ is the average molecular mass of the entire gas.
For both $\bar{d}$ and $\bar{m}$, the number density weighted averages are used.
The motivation for using this approach for diffusion coefficients is that in the atmospheres considered in this paper, molecular diffusion is especially important in a small region immediately above where H$_2$O is photodissociated and the gas composition changes rapidly from H$_2$O to H and O.
In this region, which typically extends a few hundred km in altitude, the composition of the background gas changes from being dominated by water molecules to being dominated by atomic hydrogen and oxygen, with several other molecules being abundant at different altitudes.
Therefore, I consider a more general and theoretical approach for diffusion coefficients to be more appropriate here than using measured values that are only appropriate for specific background gases.
However, the two approaches lead to very similar diffusion coefficients with values that are typically within a factor of $\sim$1.5 at the altitudes at which molecular diffusion is important.

\footnotetext{
The diameters of each chemical species was taken from the online chemical platform \url{chemicalize.com} developed by ChemAxon.  
}

In Fig.~\ref{fig:Case5}, I show an example simulation for a water vapor atmosphere of an Earth-mass planet orbiting an active solar mass star.
This is Case~5 presented in the following section. 
The strong stellar XUV field heats the gas to such a high temperature that it accelerates away from the planet in the form of a transonic Parker wind, with the point where the atmosphere becomes supersonic being also where the outflow speed exceeds the escape speed. 
The dissociation of the water molecules takes place very low in the atmosphere, at an altitude of $\sim$100~km, and the upper atmosphere is dominated by atomic hydrogen and oxygen, and their ion equivalents.

\section{Results} \label{sect:results}

\begin{figure}
\centering
\includegraphics[width=0.45\textwidth]{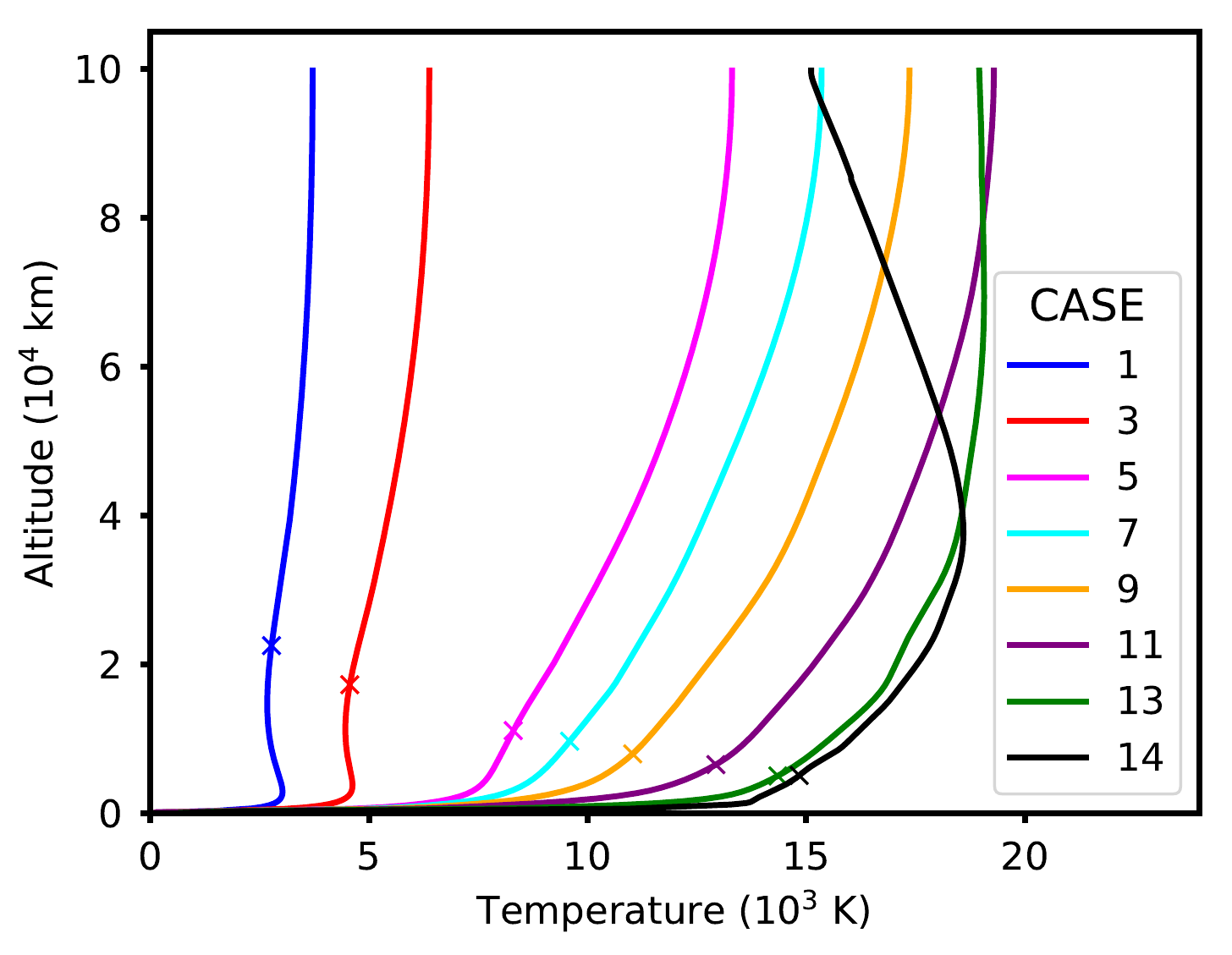}
\includegraphics[width=0.45\textwidth]{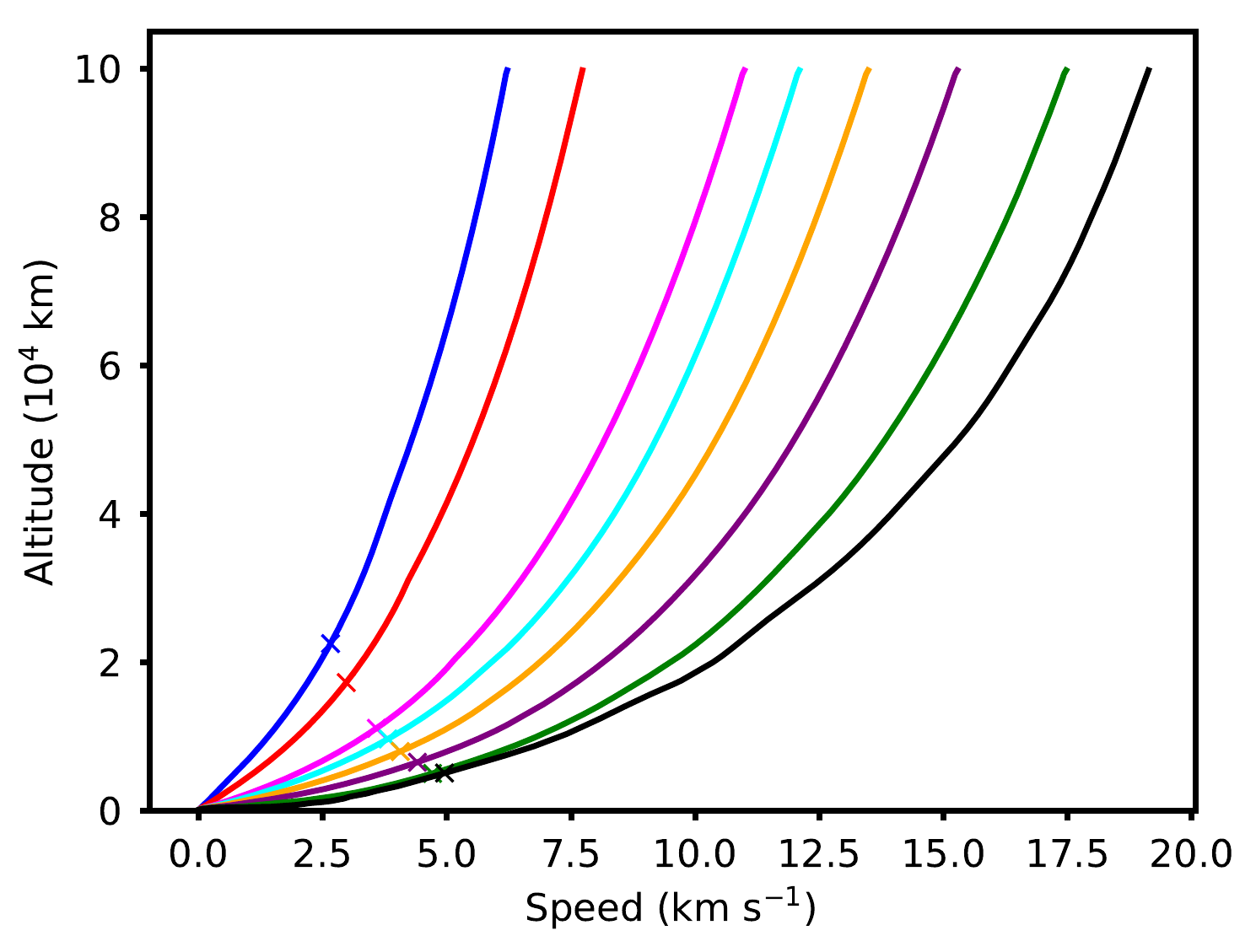}
\includegraphics[width=0.45\textwidth]{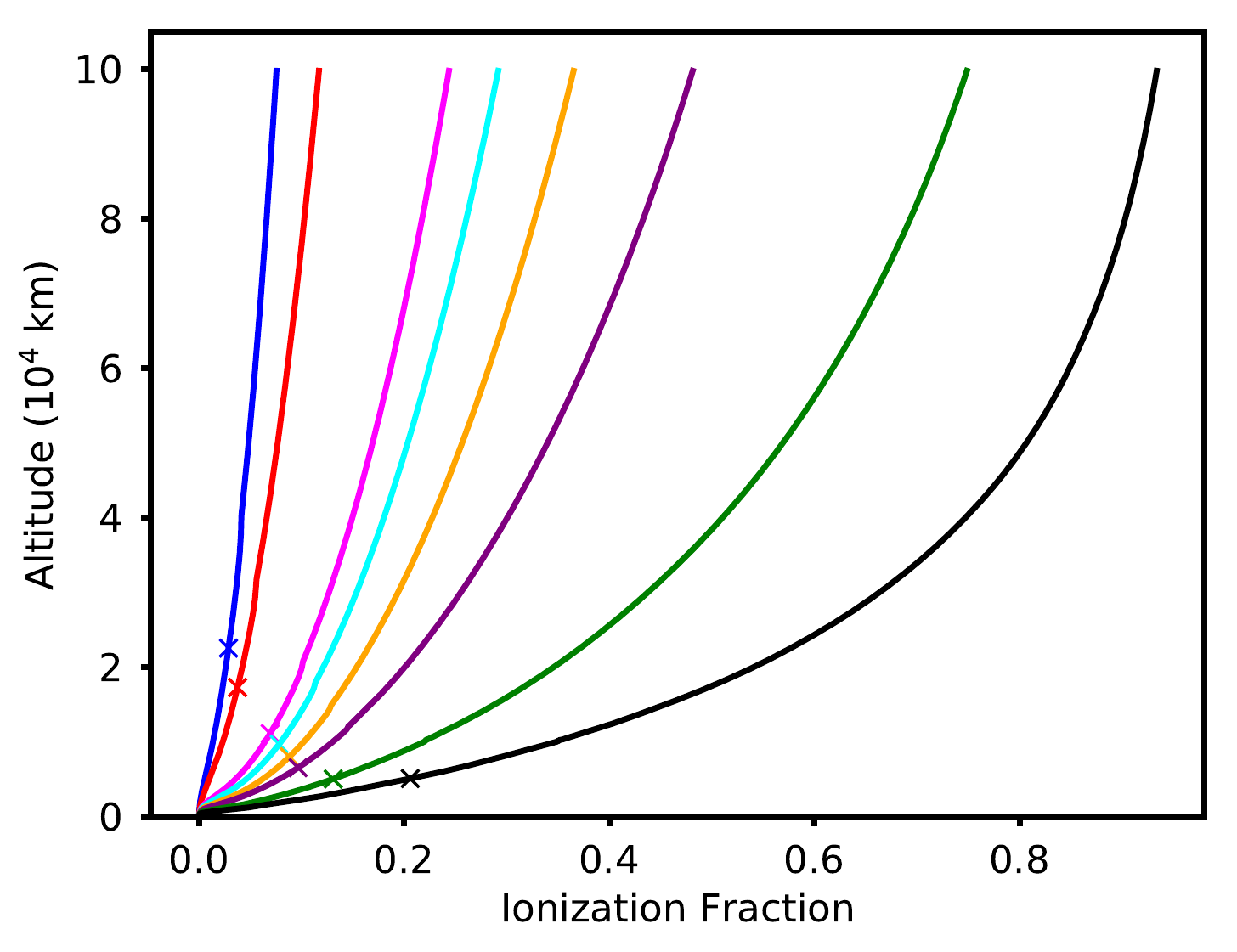}
\caption{
Vertical structures of the hydrodynamically outflowing atmospheres for several cases with different input stellar XUV fluxes, showing outflow velocity (\emph{upper-panel}), temperature (\emph{middle-panel}), and ionization fraction (\emph{lower-panel}) as functions of altitude throughout the simulation domain.
In all panels, the crosses on each line show where the atmospheres become supersonic.
}
\label{fig:hydrogrid}
\end{figure}

\begin{figure}
\centering
\includegraphics[width=0.45\textwidth]{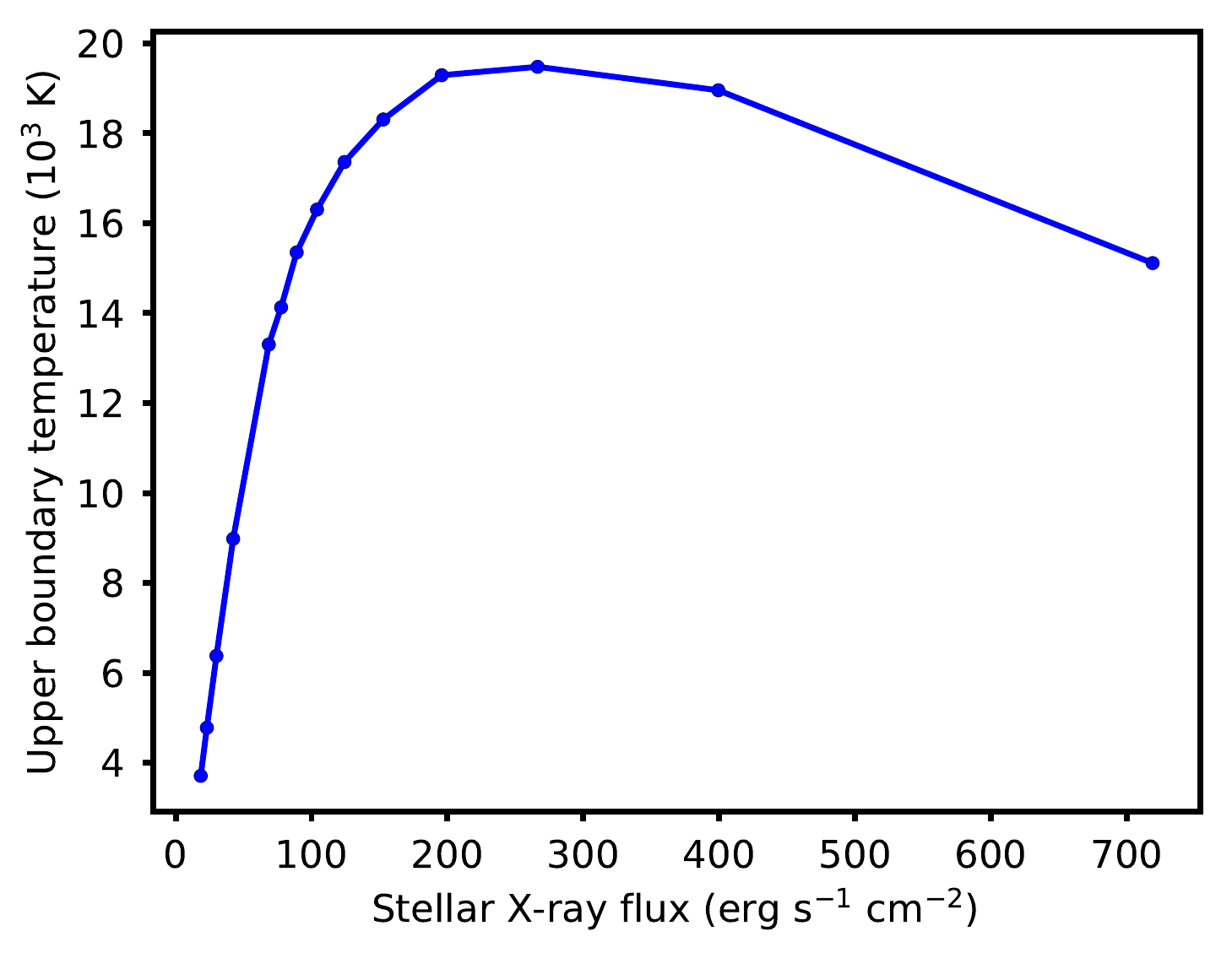}
\includegraphics[width=0.45\textwidth]{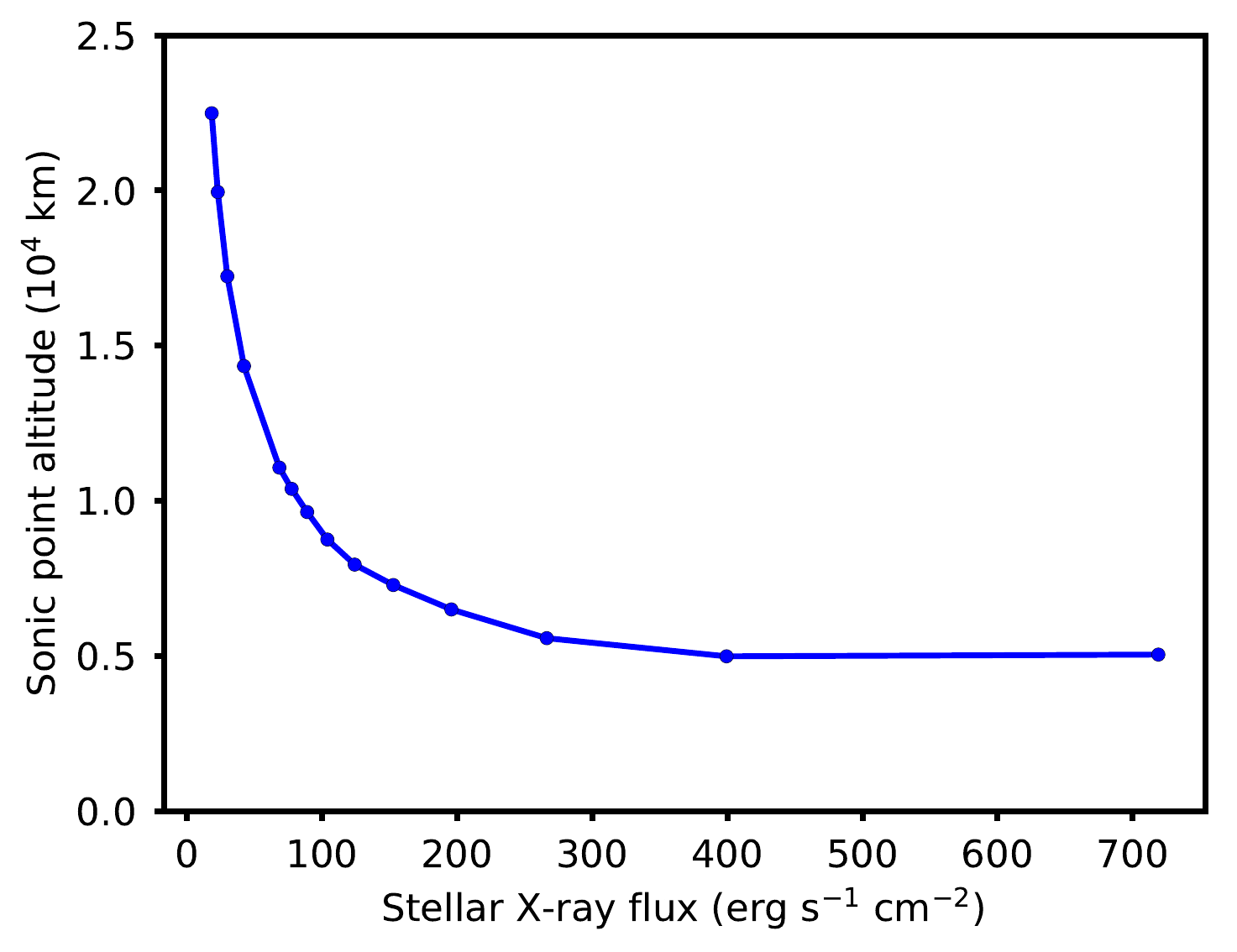}
\includegraphics[width=0.45\textwidth]{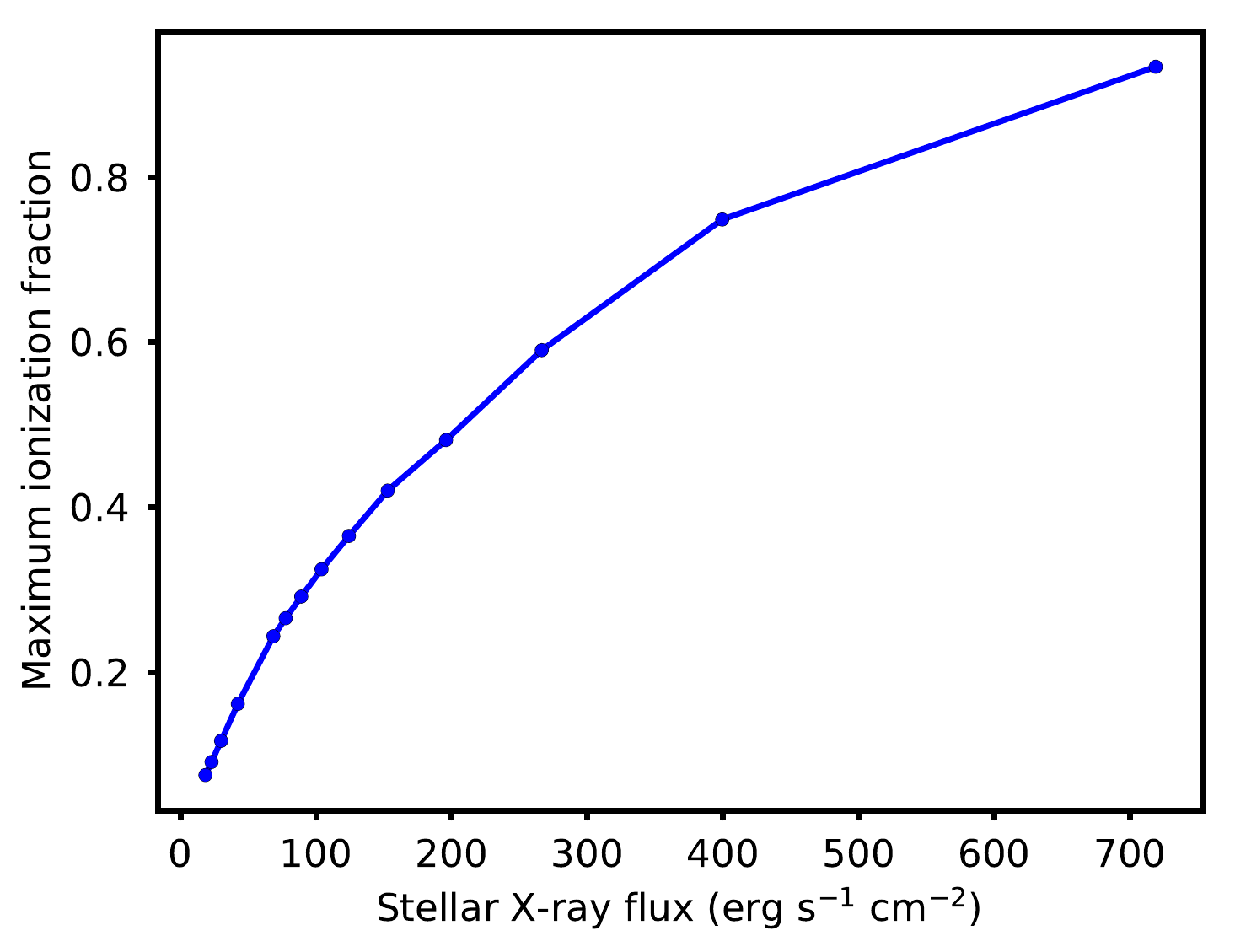}
\caption{
Several properties of the outflowing atmospheres as functions of the input stellar X-ray flux.  
}
\label{fig:hydrogridFx}
\end{figure}

\begin{figure}
\centering
\includegraphics[width=0.43\textwidth]{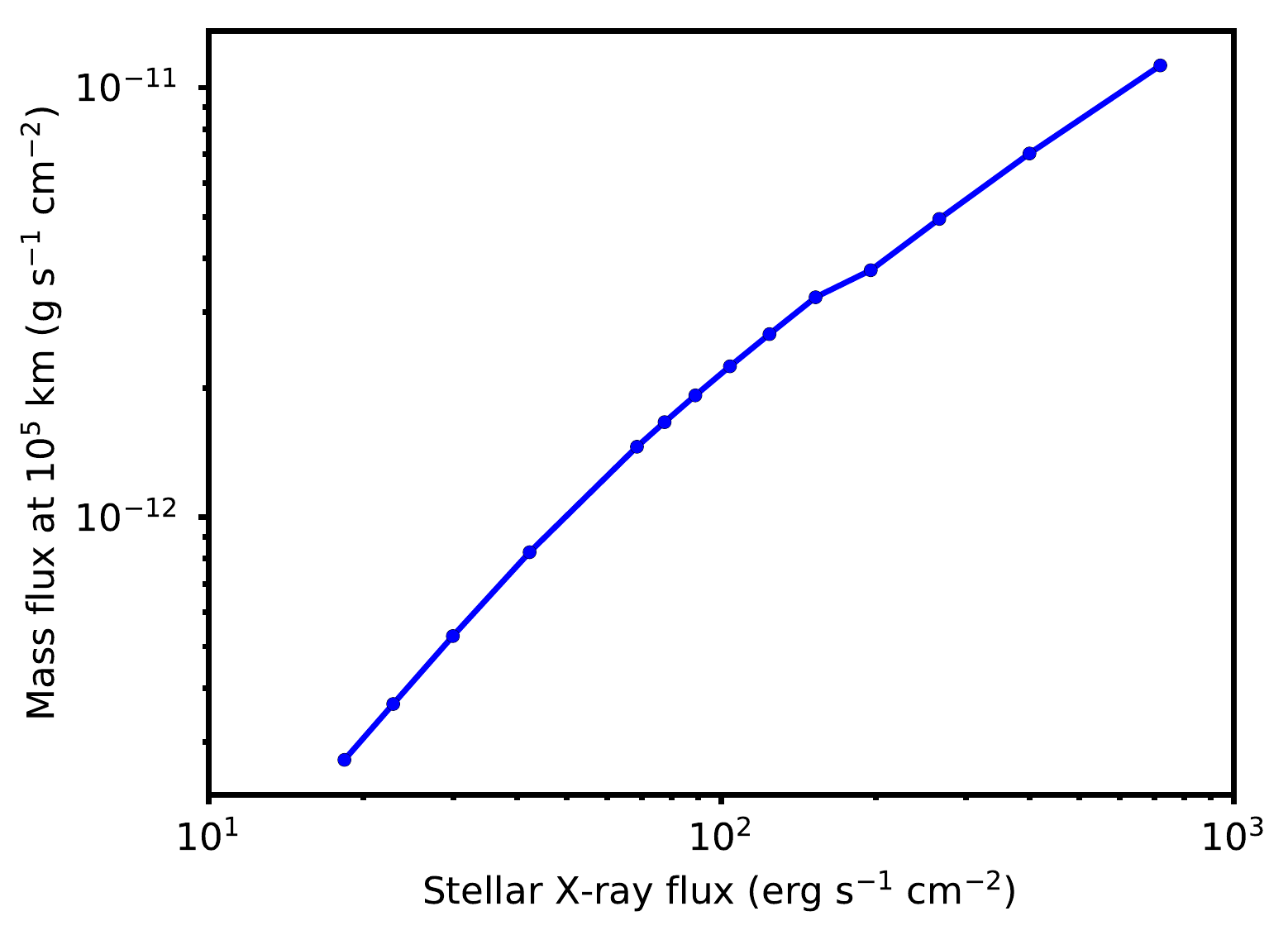}
\includegraphics[width=0.43\textwidth]{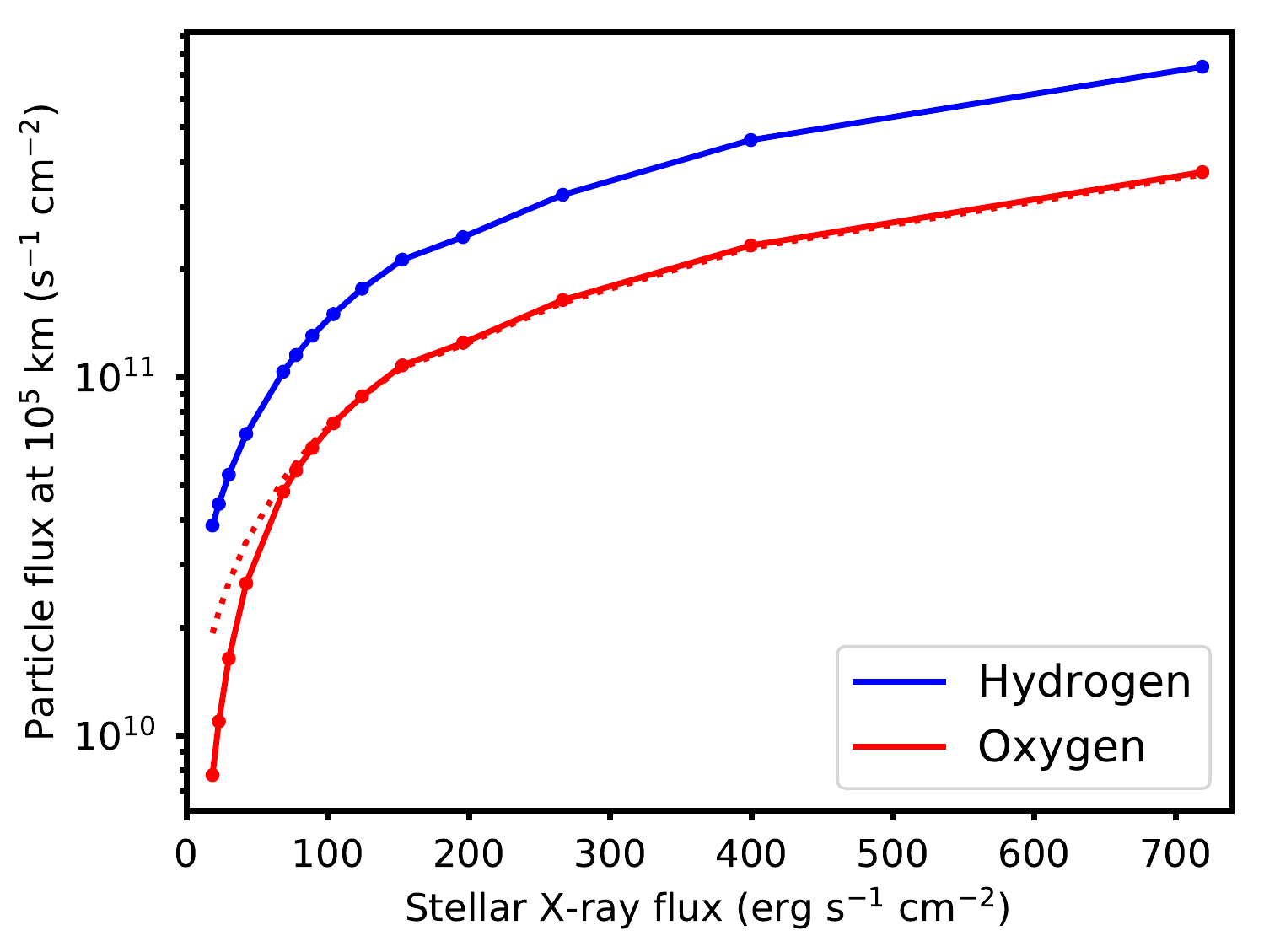}
\includegraphics[width=0.43\textwidth]{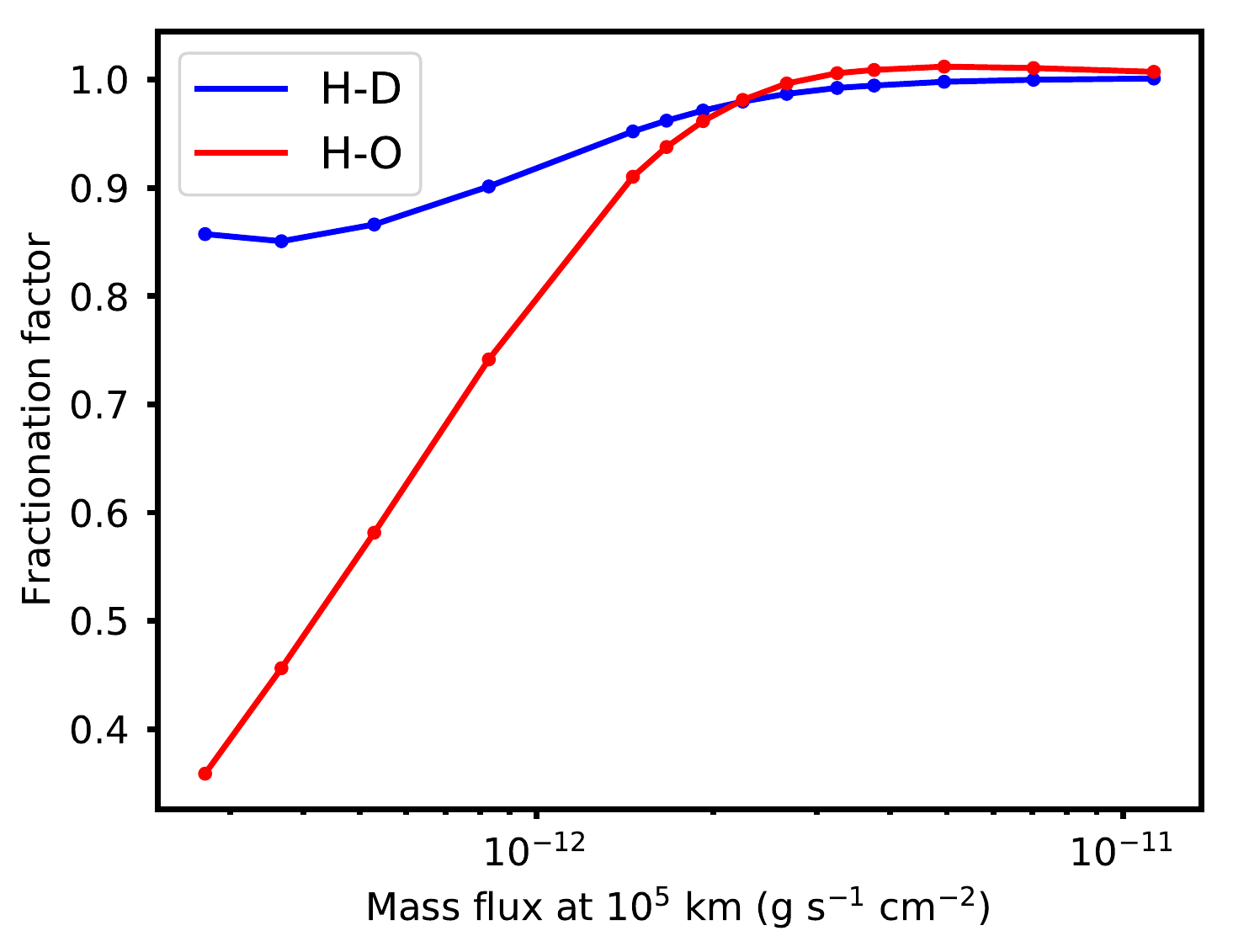}
\caption{
\emph{Upper-panel:}
outward hydrodynamic mass flux at $10^5$~km altitude as a function of input stellar X-ray flux.  
\emph{Middle-panel:}
outward hydrodynamic particle fluxes of oxygen and hydrogen atoms (including those contained in molecules and ions), with the solid lines and circles showing the simulation results and the dotted red line showing half the hydrogen flux, indicating the oxygen loss rate expected if the dissociation products of H$_2$O were lost with the same efficiency.
\emph{Lower-panel:}
fractionation factors for H and D, $f_\mathrm{H,D}$, and for H and O, $f_\mathrm{H,O}$, as a function of the mass flux. 
In all panels, the circles show the results of individual simulations 
}
\label{fig:MdotgridFx}
\end{figure}

To understand the hydrodynamic outflow of water vapor atmosphere, I calculate fourteen models for an Earth-mass planet with a fully H$_2$O atmosphere orbiting a solar mass star at 1~AU. 
The cases considered differ in the input stellar XUV spectrum calculated using the method and codes presented by \citet{Claire12}.
Specifically, I use the their solar spectra corresponding to solar ages of 4.25, 4.30, 4.35, 4.40, 4.45, 4.46, 4.47, 4.48, 4.49, 4.50, 4.51, 4.52, 4.53, and 4.54~Gyr ago.
Note however that the correspondence between age and activity presented in \citet{Claire12} is likely incorrect given that the early activity of the Sun is not known and would have depended sensitively on the Sun's initial rotation rate (\citealt{Johnstone15}; \citealt{Tu15}).
In Fig.~\ref{fig:XUVtracks}, three possible evolutionary tracks for the Sun's rotation are shown, with the grey shaded area showing the range of solar X-ray fluxes at 1~AU that I consider.

In Table.~\ref{table:grid}, I give a summary for each considered case of the input XUV spectrum fluxes and the expected age ranges that the early Sun, or any solar-mass star, would have these fluxes.
For the age ranges, I use the slow and rapid rotator tracks for X-ray emission shown in Fig.~\ref{fig:XUVtracks}.
These tracks give the expected evolutions of solar mass stars born at the 10th and 90th percentiles of the distribution of rotation rates.
Therefore, for each input spectrum, the lower limit on the age is the age at which the slow rotator reaches the corresponding activity level of the spectrum and the upper limit is the age at which the fast rotator reaches this activity level.
This is meant as an approximate measure of the expected age for a given X-ray flux and does not take into account the fact that stars may lie below 10th or above the 90th percentiles of the rotation distribution and does not take into account the large spread in X-ray luminosities for stars with a given mass, rotation rate, and luminosity (e.g. see Fig.~3 of \citealt{Reiners14}).


\begin{figure*}
\centering
\includegraphics[width=0.43\textwidth]{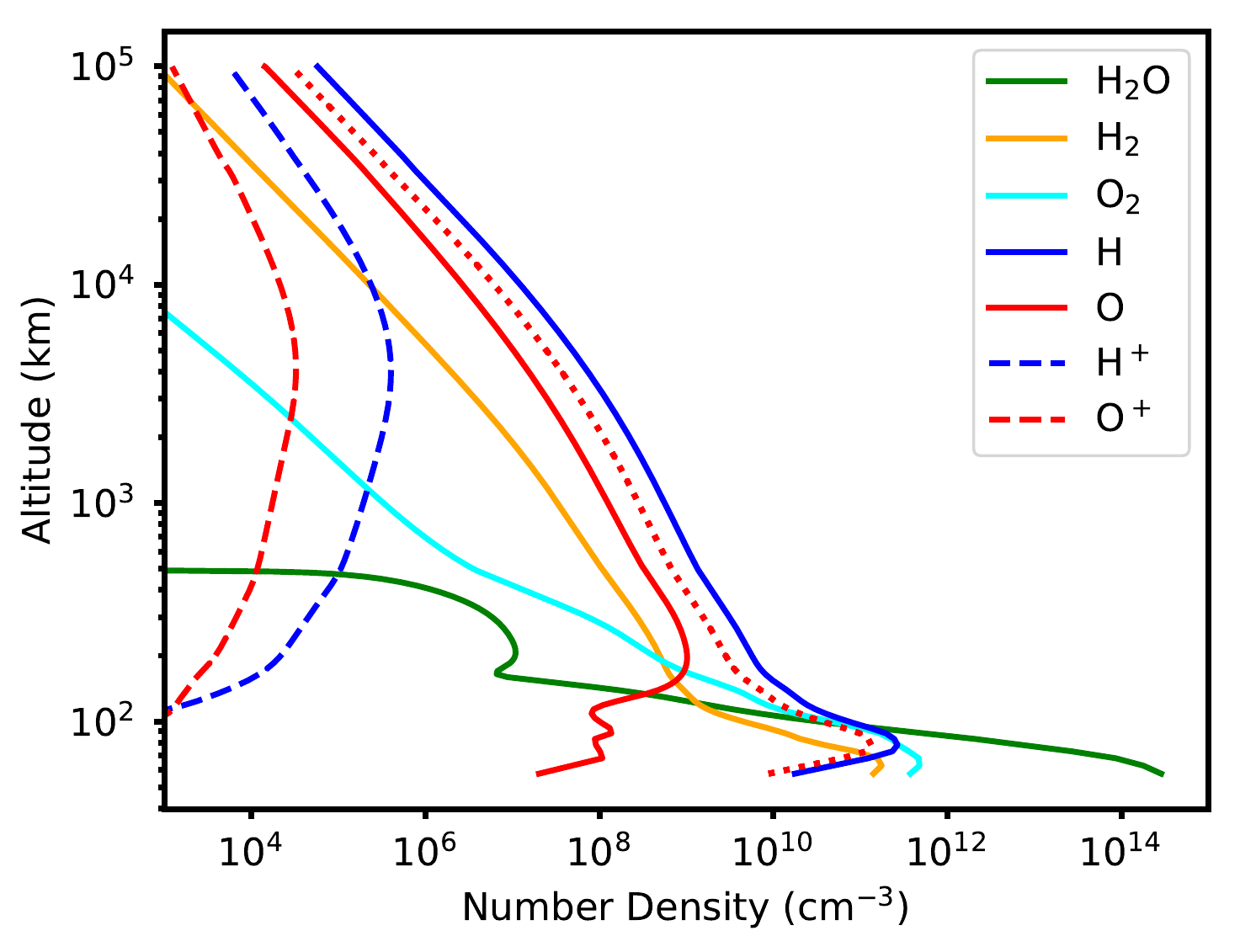}
\includegraphics[width=0.43\textwidth]{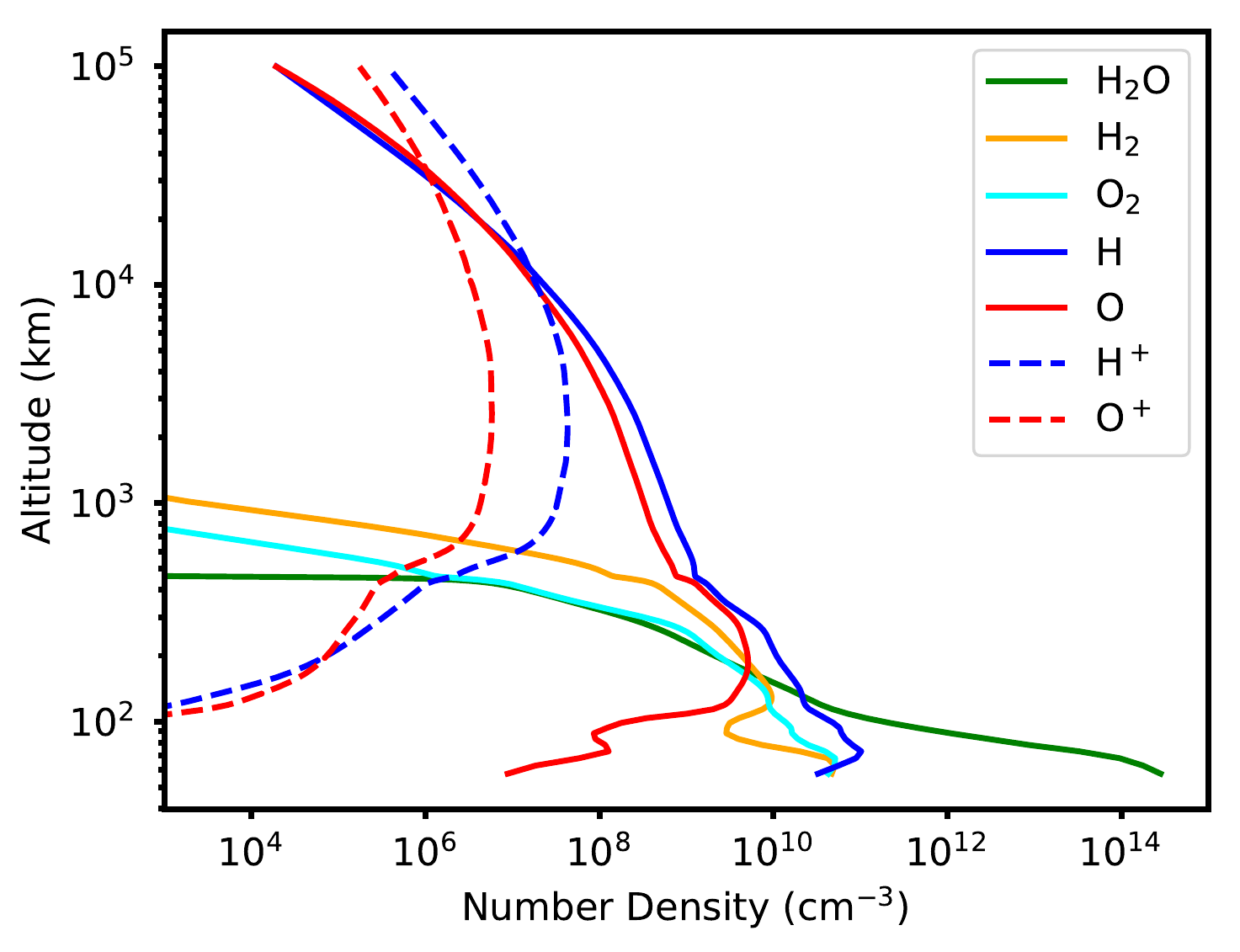}
\includegraphics[width=0.43\textwidth]{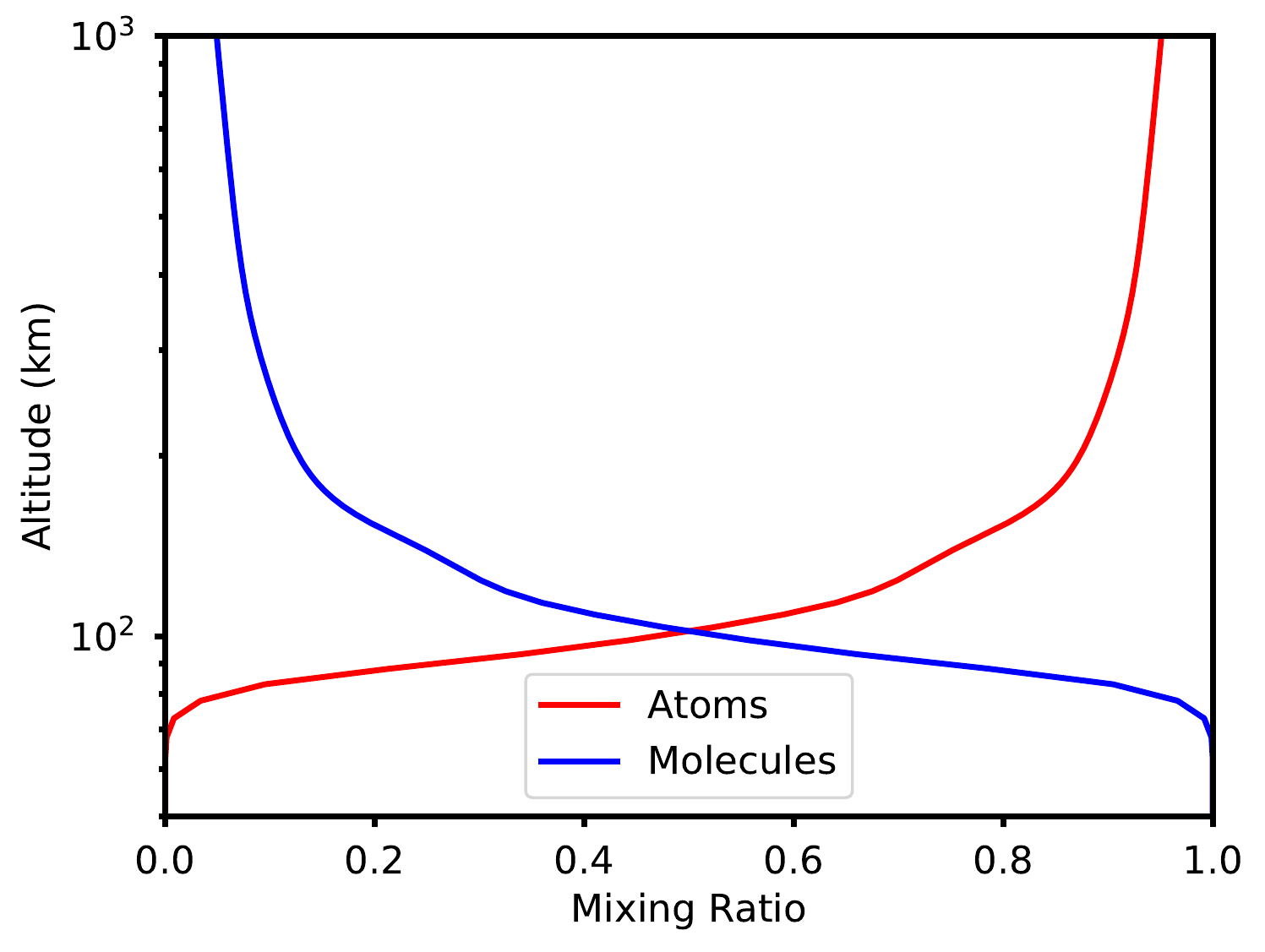}
\includegraphics[width=0.43\textwidth]{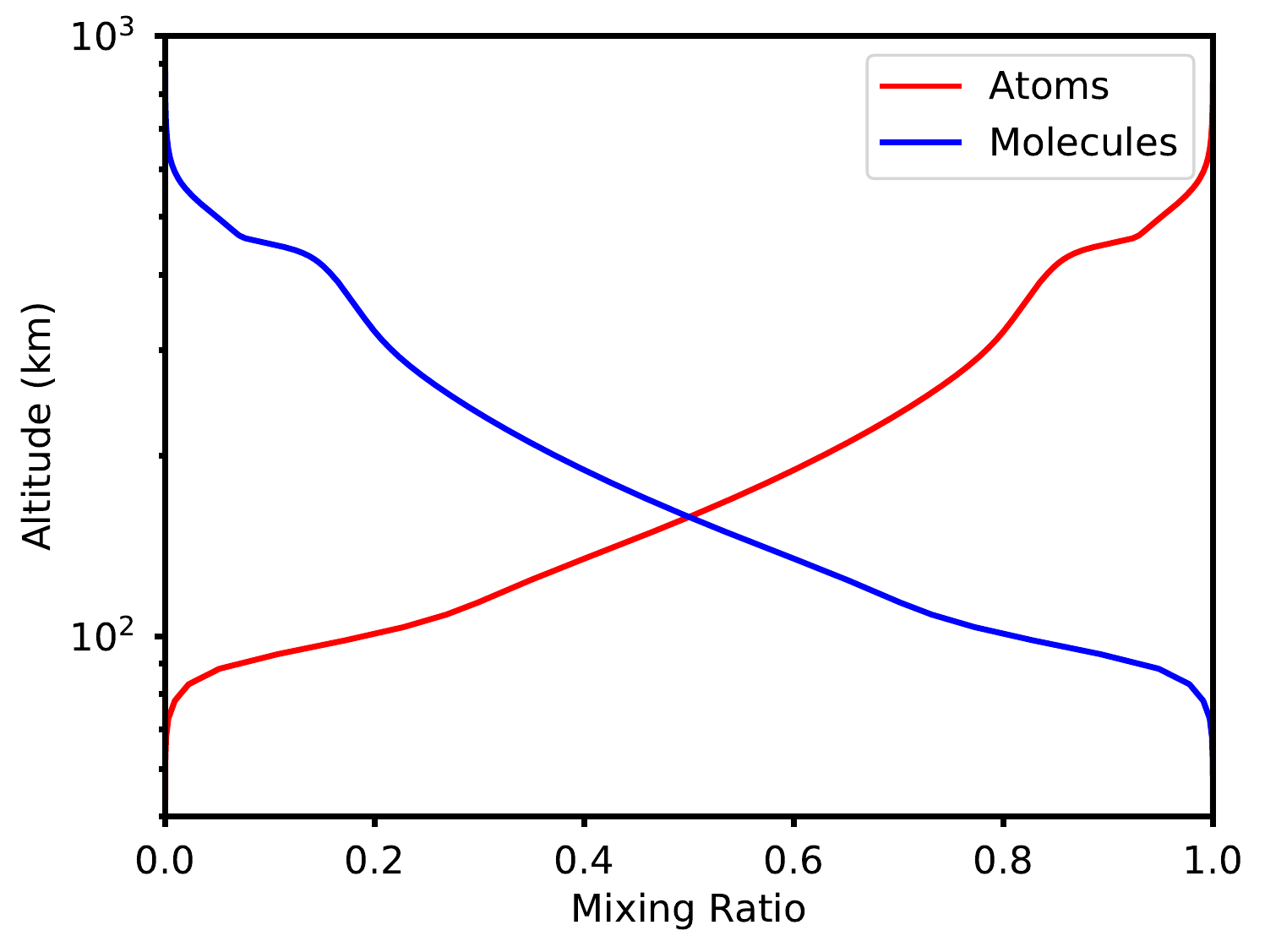}
\includegraphics[width=0.43\textwidth]{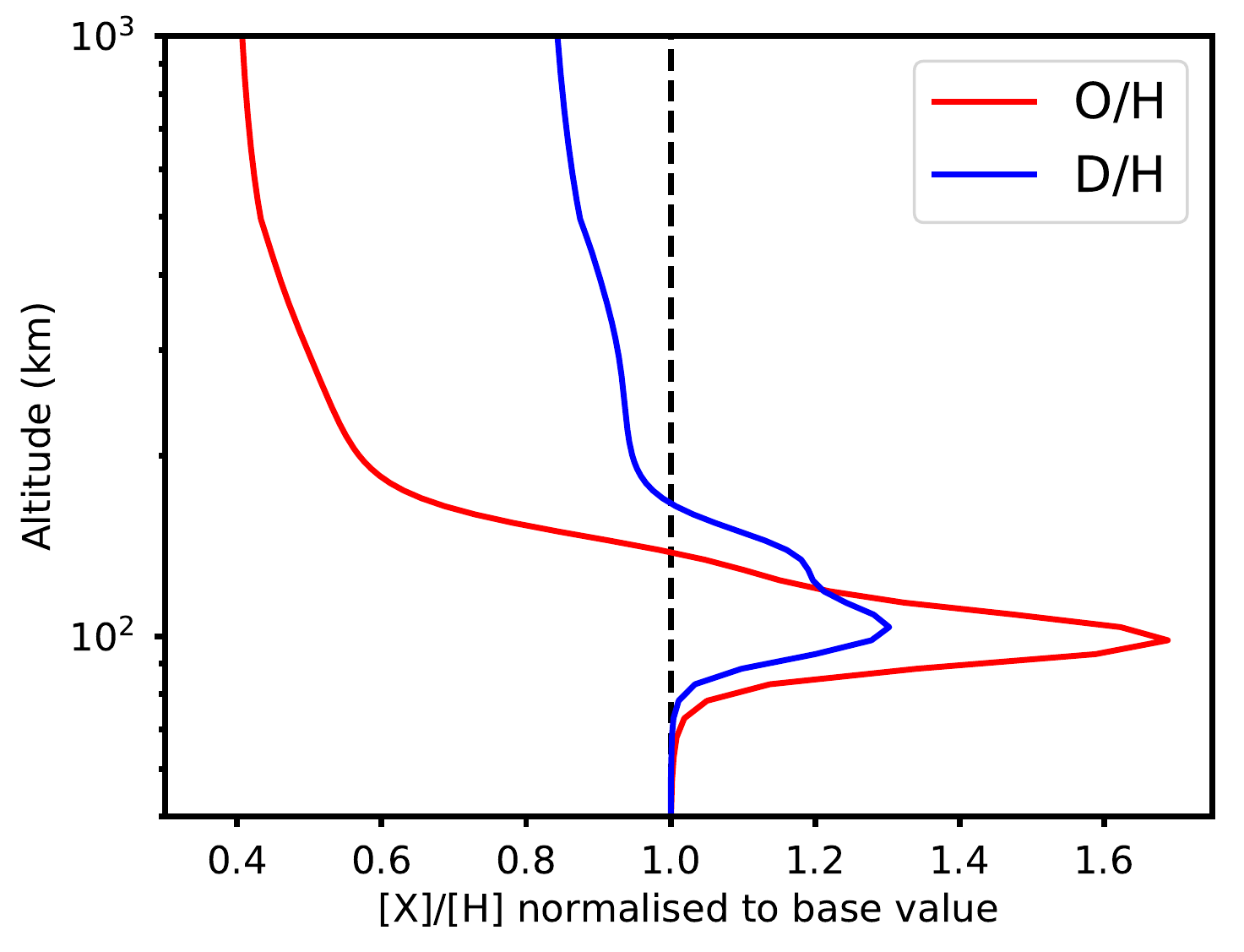}
\includegraphics[width=0.43\textwidth]{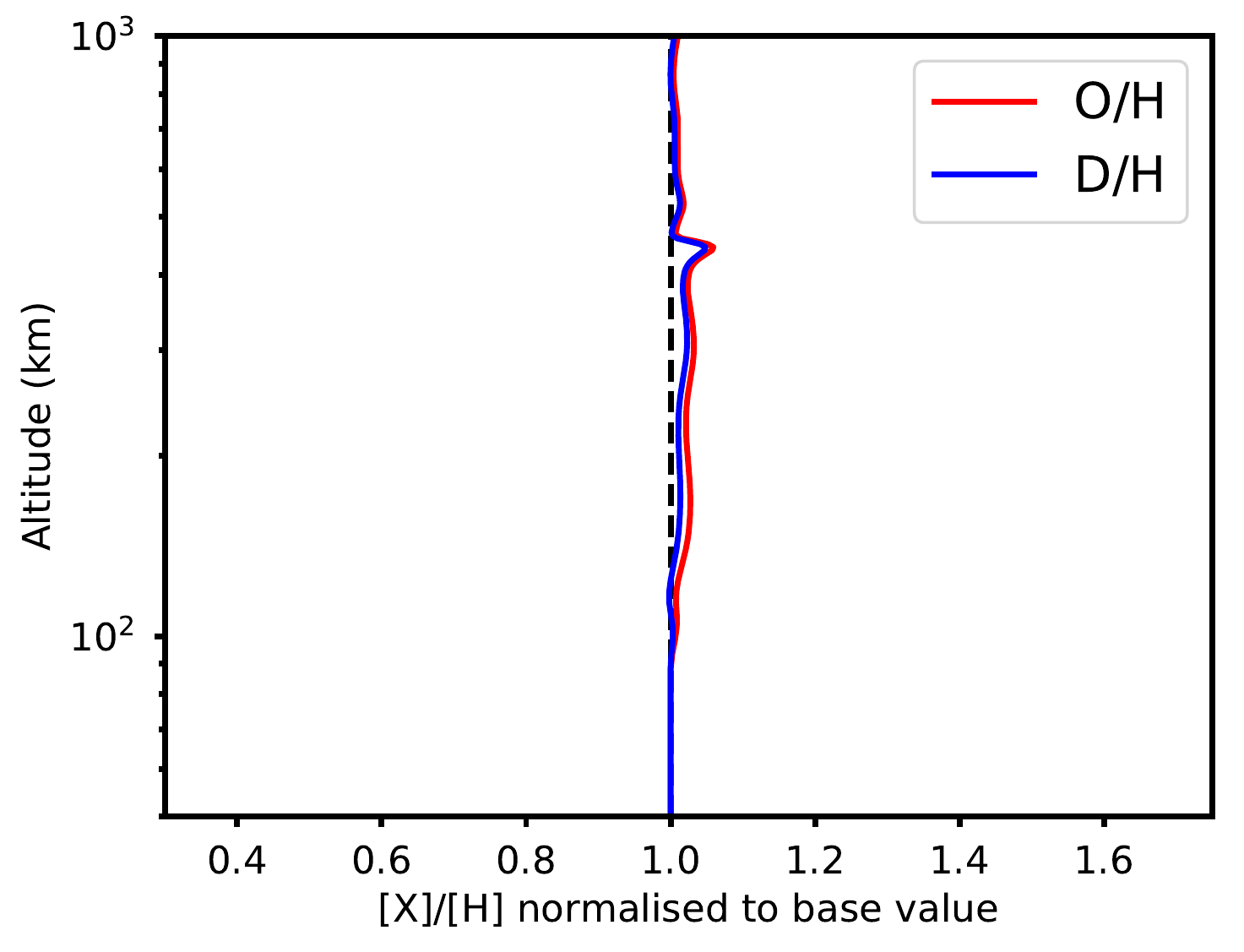}
\caption{
Chemical structures of the model atmospheres from Case~2 (\emph{left-column}) and Case~14 (\emph{right-column}).
The top row shows the densities of several neutral and ion species, the middle row shows the total molecular and atom mixing ratios, and lower row shows the D/H and O/H ratios normalised to the base values.
In the mixing ratio panel for Case~2, the dotted red line shows half the hydrogen mixing ratio, indicating the approximate mixing ratio structure that would be expected for atomic oxygen if the dissociation products of H$_2$O were lost with approximately the same efficiency. 
}
\label{fig:chem}
\end{figure*}


In Fig.~\ref{fig:hydrogrid}, the vertical structures of the gas temperature, outflow speed, and ionization fraction are shown for several of the cases calculated.
In all cases, the absorption of the stellar XUV spectrum heats the gas to very large temperatures, causing the atmosphere to flow away from the planet hydrodynamically in the form of a transonic wind. 
At altitudes of $\sim$5000-20,000~km, the outflow speeds exceeds the escape velocity; this point is also the sonic point as required in a thermal pressure driven Parker wind and is marked by the crosses in Fig.~\ref{fig:hydrogrid}. 
In no case is the exobase found within the simulation domain.

The dependence on the input X-ray flux of the maximum gas temperature, the sonic point altitude, and the maximum ionization fraction are shown in Fig.~\ref{fig:hydrogridFx}.
In the higher activity cases, the atmospheres are heated to higher temperatures, accelerated to larger outflow speeds, and become more ionized.
In the highest activity cases, the temperature at the upper boundary of the simulation stops increasing with increasing activity and instead decreases due to adiabatic cooling, which can cause a negative temperature gradient in hydrodynamic outflows (\citealt{Tian08}; see also Fig.~1 of \citealt{Johnstone15letter}).
This can be best seen in the upper panel of Fig.~\ref{fig:hydrogridFx}, where the maximum temperature of $\sim$20,000~K is reached at an input X-ray flux of $\sim$250~erg~s$^{-1}$~cm$^{-2}$.
The sonic point altitude, shown in the lower panel of Fig.~\ref{fig:hydrogridFx}, gets closer to the planet as the input XUV flux increases due to the more rapid acceleration of the wind, but at very high activity cases appears to become approximately independent of activity with a value of $\sim$500~km.
The ionization fraction at the upper boundary continues to increase with increasing activity, even at very high activity level, and in the most active case considered is 93\% meaning that as it flows away from the planet, the gas is almost entirely ionized.

In Fig.~\ref{fig:MdotgridFx}, I show the outward hydrodynamic mass flux as a function of input stellar X-ray flux.
As expected, higher stellar activity cases have significantly higher atmospheric losses.
In and near the range of input X-ray fluxes considered, the mass flux at the upper boundary of the simulation domain (at an altitude of $10^5$~km) can be described by  
\begin{equation} \label{eqn:Mdot}
\log F_\mathrm{mass} = 
-0.239 \left( \log F_\mathrm{X} \right)^2 
+ 1.982 \log  F_\mathrm{X}
- 3.681 ,
\end{equation}
where the units of $F_\mathrm{X}$ are erg~s$^{-1}$~cm$^{-2}$ and $F_\mathrm{mass}$ are \mbox{10$^{-11}$~g~s$^{-1}$~cm$^{-2}$}. 
As a note of caution, it is not trivial to estimate total atmospheric mass loss rates from these fluxes and this problem is discussed in more detail in Section~\ref{sect:evo}.
Since I calculate 1D hydrodynamic models for the direction pointing in the direction of the star, simply multiplying the mass flux by \mbox{$4 \pi R^2$}, where $R$ is the Earth's radius plus $10^5$~km, would overestimate the mass loss rates by a factor of a few.
The outward particle fluxes for hydrogen and oxygen and the hydrogen-oxygen and hydrogen-deuterium fractionation factors are also shown in Fig.~\ref{fig:MdotgridFx} and discussed below.

For Case~2 and Case~14, representing moderate and high activity cases, information about the chemical structures of the upper atmospheres is shown in Fig.~\ref{fig:chem} for five neutral and two ion species.
At the base of the simulation, the gas is composed entirely of H$_2$O, which is then quickly photodissociated, creating several molecular and atomic species including H$_2$, O$_2$, H, and O.
At slightly higher altitudes, the H$_2$ and O$_2$ are then also photodissociated and the gas becomes completely dominated by H and O. 
Between altitudes of approximately 100 and 1000~km, H$^+$ and O$^+$ start to become important, and in the high activity case,  H$^+$, O$^+$, and free electrons are the most abundant species by the upper boundary of the simulation domain.  
It is important to note that at no point in the atmosphere is the gas dominated by atomic hydrogen: while H has the highest number density throughout most of the atmosphere, the heavier O has the highest mass density of all species.  
In Fig.~\ref{fig:chem}, I also show the total mixing ratios of atoms and molecules in both simulations showing where in the atmosphere most of the dissociation of molecules takes place.


\begin{figure*}
\centering
\includegraphics[width=0.45\textwidth]{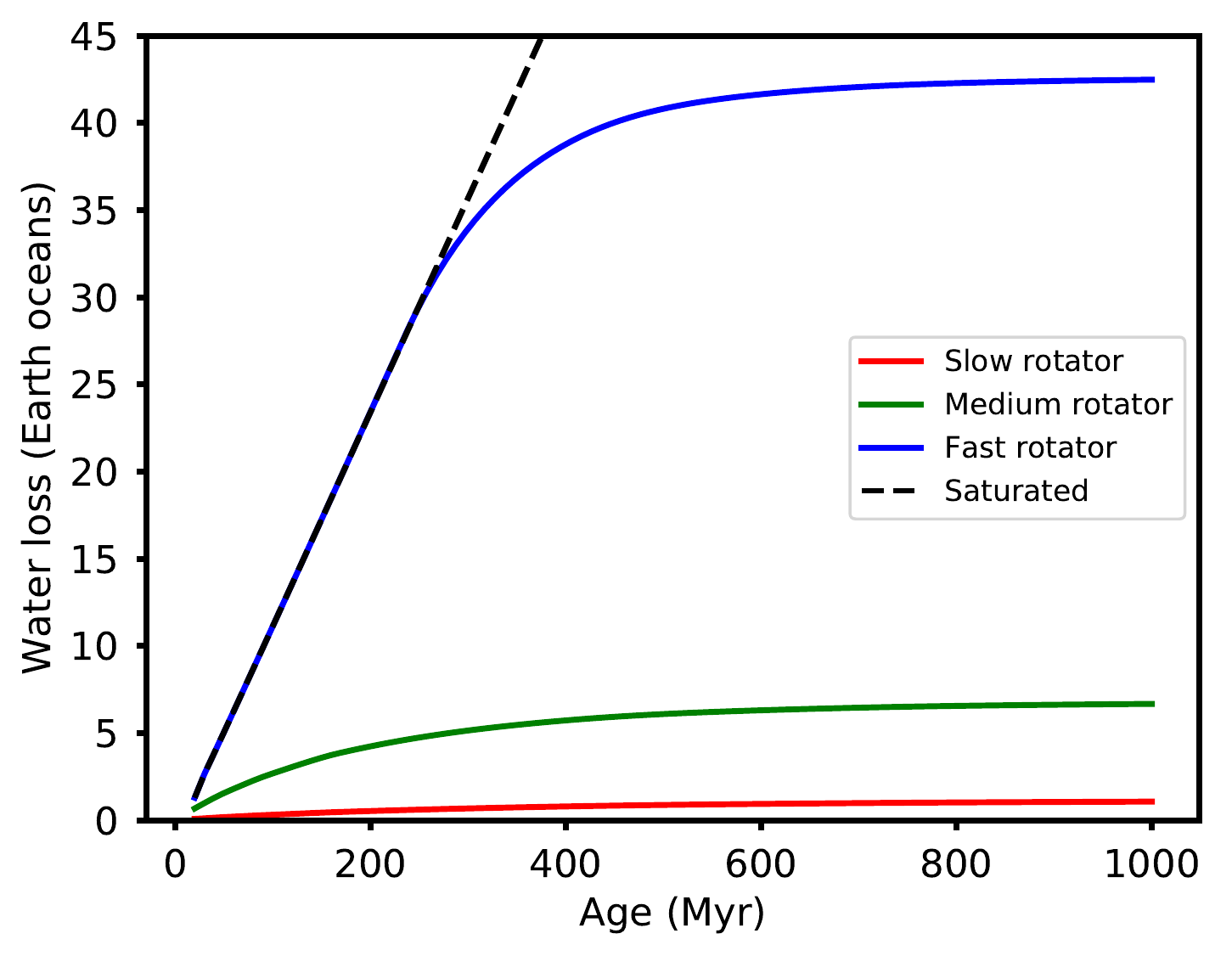}
\includegraphics[width=0.45\textwidth]{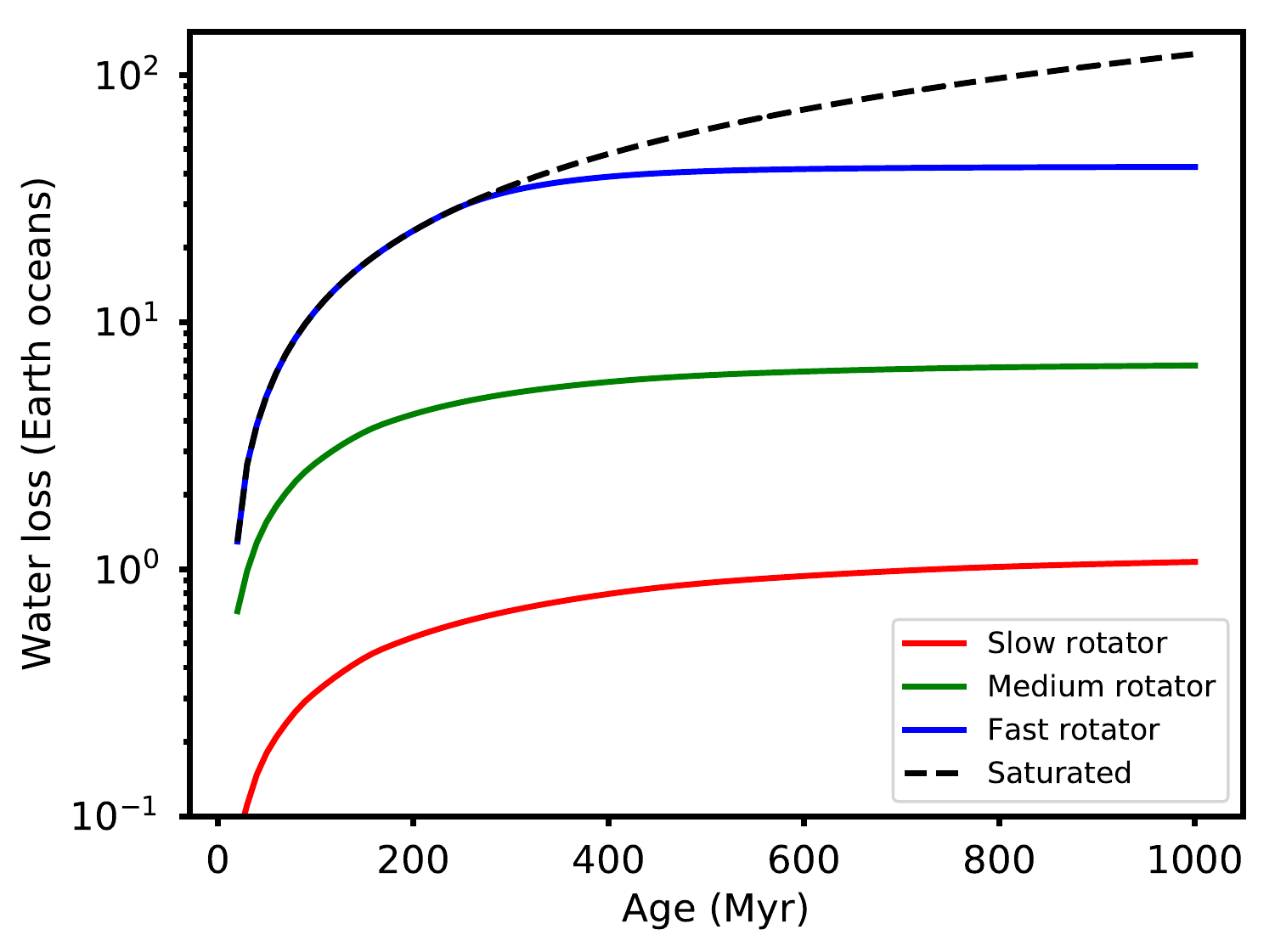}
\caption{
Cumulative evolution of water lost from a water vapor atmosphere of an Earth-mass planet orbiting a solar mass star at 1~AU.
The red, green, and blue lines represent the cases of initially slow, medium, and fast rotating stars respectively, and the dashed line represents the case of a star that remains at the saturation threshold for the entire first billion years.
The black line represents what might be expected for a planet orbiting in the habitable zone of a fully-convective M star.
The left and right panels differ only in the y-axis scale. 
}
\label{fig:MlossGrid}
\end{figure*}



\begin{figure}
\centering
\includegraphics[width=0.45\textwidth]{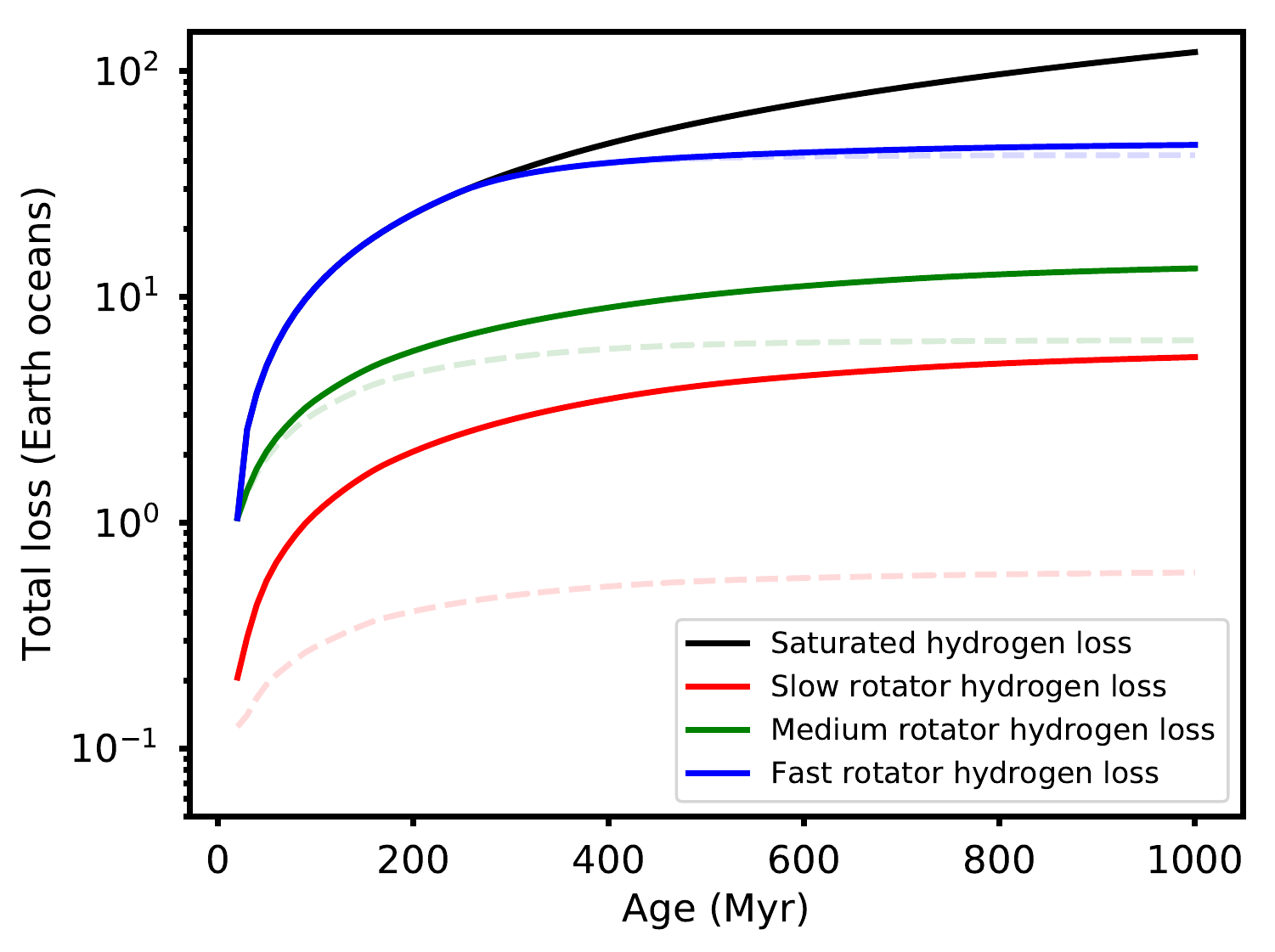}
\includegraphics[width=0.45\textwidth]{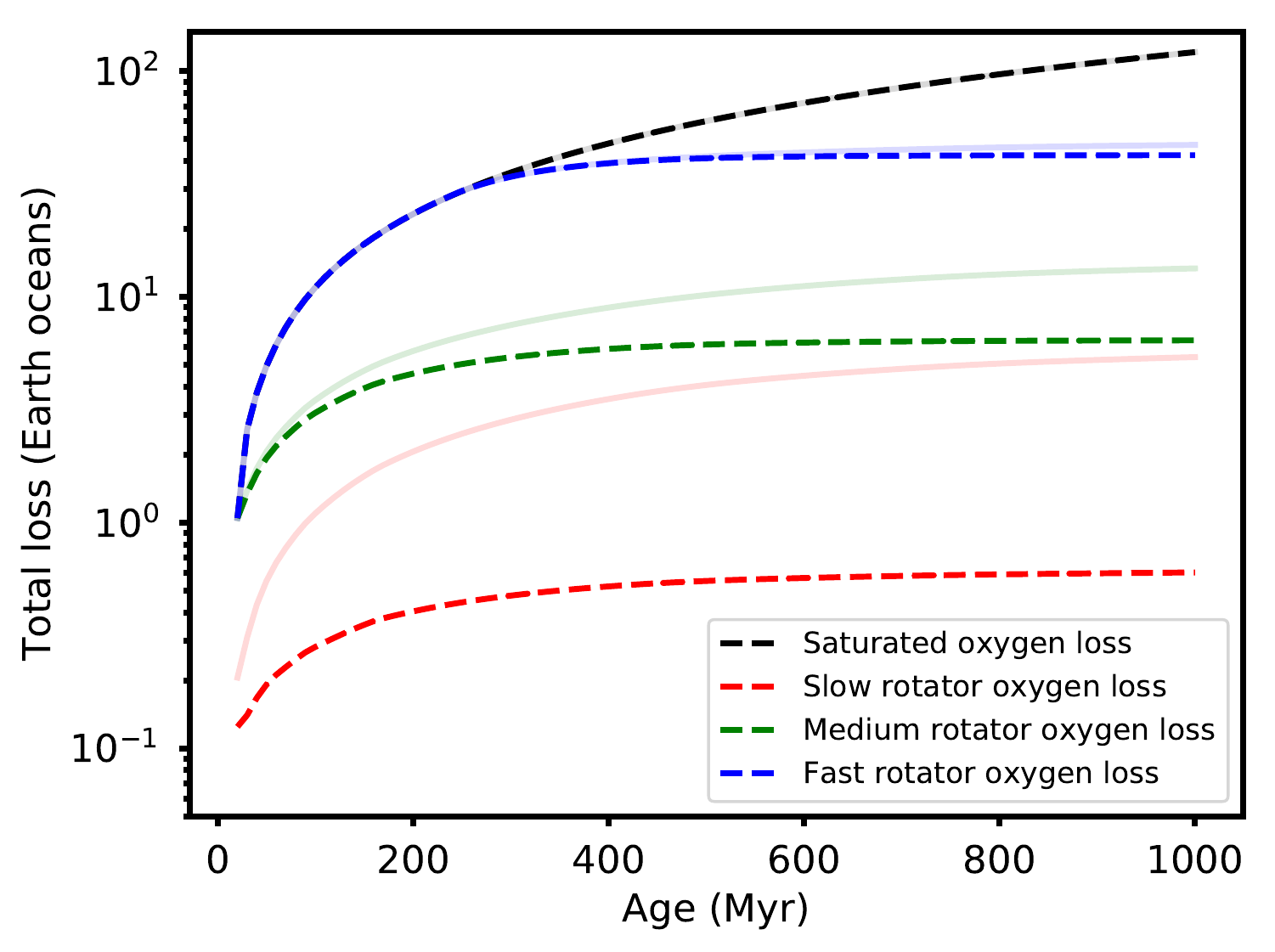}
\includegraphics[width=0.45\textwidth]{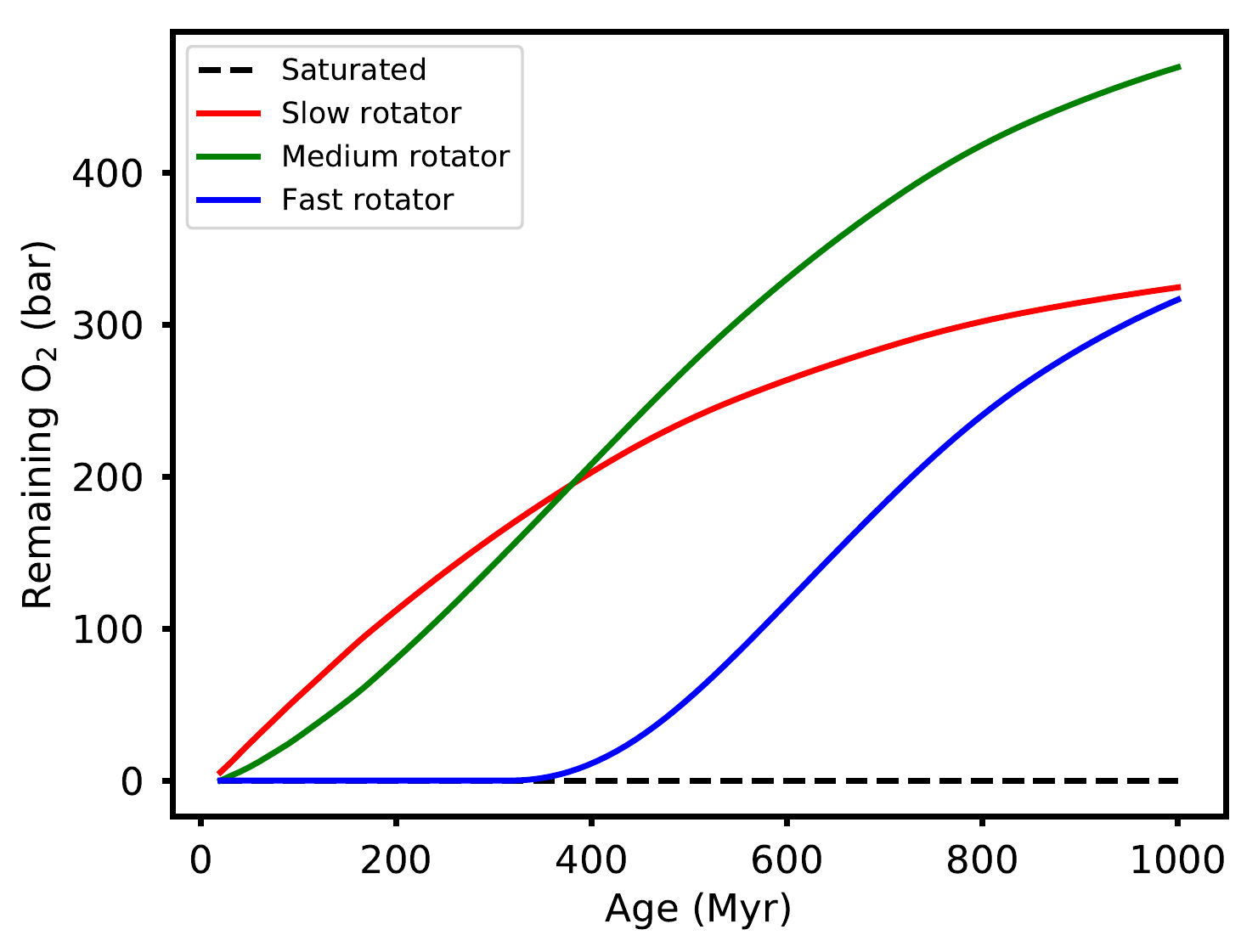}
\caption{
Cumulative evolution of hydrogen (\emph{upper-panel}) and oxygen (\emph{middle-panel}) losses for the evolutionary cases shown in Fig.~\ref{fig:MlossGrid}, with the amount of oxygen that builds up in the atmosphere as a result of the different loss efficiencies of hydrogen and oxygen (\emph{lower-panel}).
In the lower panel, I assume 1 Earth ocean is 270~bar.
}
\label{fig:NlossGrid}
\end{figure}


In the upper-left panel of Fig.~\ref{fig:chem}, the dotted red line shows half of the hydrogen mixing ratio, indicating approximately the values expected if the dissociation products of H$_2$O were lost equally for Case~2. 
The lower values for O relative to this line in the upper thermosphere show that H is lost more efficiently than O, as expected given its lower mass.
The reason for the different loss efficiencies of H, D, and O in the atmosphere is that in the region where the H$_2$O is photodissociated, molecular diffusion is very important and separates the particles by mass.
This is shown more effectively by the red line in the lower-left panel of Fig.~\ref{fig:chem}, which shows the ratio of the densities of oxygen atoms to hydrogen atoms (including atoms held in molecules and ions) divided by the base value.
In this region around an altitude of $\sim$100~km, initially H becomes less abundant relative to O because it is more rapidly transported upwards by molecular diffusion and then becomes more abundant for the same reason.
A similar trend is seen in the D/H ratio, shown as the blue line in the lower-left panel of Fig.~\ref{fig:chem}, though the separation by mass is much less significant since the masses of D and H are much more similar.

This mass fractionation is much lower than we would expect for a fully hydrostatic atmosphere because molecular diffusion is only able to separate the species by mass in a small range of altitudes above the homopause.
The homopause is defined as the altitude at which the effect of molecular diffusion exceeds the effect of eddy diffusion, and below the homopause, eddy diffusion forces all the species that are not quickly created and destroyed by chemical reactions to have mixing ratios that are uniform with altitude.
This is typically at the base of the thermosphere since this is where the rapid increase in temperature due to XUV heating causes a rapid increase in the molecular diffusion coefficients.
Above the homopause, molecular diffusion separates the species by mass; however, in a hydrodynamically outflowing atmosphere, the hydrodynamic advection of the gas also has the effect of making the mixing ratios of each species uniform with altitude, and since this is the dominant effect in the upper thermosphere, very little mass separation takes place. 
The mass fractionation is quite small because the distance between the homopause and where the advection of species dominates is very small such that molecular diffusion does not have much of an opportunity to separate the species by mass.
This is consistent with the results presented by \citet{Gillmann09} for mass fractionation of noble gases in the atmosphere of early Venus.

As can be seen in the lower right panel of Fig.~\ref{fig:chem}, for our highest activity case, almost no mass fractionation takes place.
This is due to the much stronger hydrodynamic escape that takes place in this case. 
The fractionation factor between the two escaping species, $f_{i,j}$, is defined by \mbox{$f_{i,j} = ( F_j / F_i ) (N_i / N_j)$}, where $F_i$ and $N_i$ are the escape flux and total atmospheric abundance of the $i$th particle (\citealt{Mandt09}).
It is reasonable to assume that the term \mbox{$F_j/F_i$} is equal to the ratio of the densities (\mbox{$=n_j/n_i$}) of the two particles at the upper boundary of the simulation and that \mbox{$N_i/N_j$} is equal to the ratio (\mbox{$=n_i/n_j$}) at the base of the simulation, where in both cases the densities include also atoms contained within molecules.
The fractionation factors between O and H, $f_\mathrm{H,O}$, and between D and H, $f_\mathrm{H,D}$, as functions of the input X-ray flux are shown in Fig.~\ref{fig:MdotgridFx}.
The value for $f_\mathrm{H,O}$ is 0.4 for the lowest activity case and quickly increases to 1.0 as the input XUV flux increases.
Similarly, the value for $f_\mathrm{H,D}$ increases from $\sim$0.9 to 1.0. 
Therefore, for extreme hydrodynamic escape driven by very active stars, no mass fractionation takes place, leading to no oxygen accumulation and no increase in the D/H ratio.

\section{Evolution of losses and accumulation of O$_2$} \label{sect:evo}

In this section, I study the evolution of losses from an Earth-mass planet orbiting at 1~AU around a solar mass star with a fully water vapor atmosphere between ages of 10~Myr to 1~Gyr.
For all calculations, I assume that there is an infinite reservoir of water vapor available to the atmosphere to feed these losses.
To calculate the mass loss rates, I use Eqn.~\ref{eqn:Mdot} to get the outward mass flux as a function of the input stellar XUV flux.
These fluxes are calculated using the stellar activity evolution tracks calculated by \citet{Tu15}.
However, care must be taken to ensure that the mass loss rates are not overestimated since the simulations from which Eqn.~\ref{eqn:Mdot} is derived give the mass flux in the direction of the star, which is likely a factor of a few higher than the average mass flux in all directions.
As in \citet{Erkaev13}, I assume that the loss takes place over 3$\pi$ steradians, which is based on the assumption that losses do not take place where the atmosphere is not irradiated by the star due to the planet's shadow.
As in \citet{Johnstone15letter}, I then average the incoming stellar XUV energy over this 3$\pi$ steradians, which is implemented in Eqn.~\ref{eqn:Mdot} by inputting the \mbox{$F_\mathrm{XUV}/3$} in place of the XUV flux and then calculating the mass loss rate as \mbox{$\dot{M}_\mathrm{at} = 3 \pi R^2 F_\mathrm{mass}$}, where $R$ is the radius of the upper boundary of the simulation domain.
Although this is a simple way to calculate the total atmospheric losses from 1D simulations, it has a good advantage that it makes sense from an energy conservation point of view if a zero zenith angle is assumed in the simulations.
The energy available to drive atmospheric losses is no greater than the total energy input into the atmosphere and the XUV energy that does not directly drive losses is only lost because of real atmospheric processes that decrease how efficiently the absorbed XUV energy is converted into heat and how much of that heat drives losses, as opposed to energy being lost due to geometric assumptions.
To get $f_\mathrm{H,O}$ for this mass flux, I use a simple polynomial interpolation using the relation between the mass flux and $f_\mathrm{H,O}$ shown in the lower panel of Fig.~\ref{fig:MdotgridFx}.
The loss rates for oxygen and hydrogen atoms, $\dot{N}_\mathrm{O}$ and $\dot{N}_\mathrm{H}$, are given by \mbox{$\dot{N}_\mathrm{O} = f_\mathrm{H,O} \dot{M}_\mathrm{at} / (f_\mathrm{H,O} m_\mathrm{O} + 2 m_\mathrm{H} )$} and \mbox{$\dot{N}_\mathrm{H} = 2 \dot{M}_\mathrm{at} / (f_\mathrm{H,O} m_\mathrm{O} + 2 m_\mathrm{H} )$}, where $m_\mathrm{O}$ and $m_\mathrm{H}$ are the masses of oxygen and hydrogen atoms.

In Fig.~\ref{fig:MlossGrid}, the total amount of atmosphere lost as a function of age in the first Gyr is shown for the cases of the slow, medium, and fast stellar rotator tracks, and for the case of a star that remains at the saturation threshold for the first billion years (dashed black line).
For the slow stellar rotator case, the total atmospheric mass loss is $\sim$1~Earth oceans (\mbox{$1.4 \times 10^{24}$~g}).
For the fast stellar rotator case, the total atmospheric mass loss is $\sim$40~Earth oceans, which is much larger since the initially rapidly rotating star remains highly active for much longer.
Due to the decay in activity after approximately 300~Myr, this is much less than the 120~Earth oceans lost if the star remains at the saturation threshold for the entire first Gyr as might be the case for a low mass M star.
In both the first 300~Myr for the fast rotator case and in the constantly saturated case, the rapid hydrodynamic escape means that the planet is unlikely to be able to hold onto any water vapor atmosphere, and water outgassed from the interior or delivered to the planet from external sources will be lost very rapidly. 
A water vapor atmosphere could survive if the planet is formed with a significant fraction of its mass as water and therefore a sufficiently large enough reservoir of water if available to compensate for the losses to space.
Also, a water vapor atmosphere could form by outgassing after the star's activity has decayed to more moderate levels.

The total losses of hydrogen and oxygen in Earth ocean equivalents for each stellar rotator case is shown in Fig.~\ref{fig:NlossGrid}.
I define an Earth ocean equivalent for a given atom to be the number of those atoms in one Earth ocean.
The differences between the dashed and solid lines shown in the upper and middle panels is due to the different loss efficiencies of H and O. 
The resulting build-up of O$_2$ molecules due to the slightly lower loss rates of O is shown in the lower panel of Fig.~\ref{fig:NlossGrid}.
For the fast rotator case, no O$_2$ is able to build-up in the atmosphere for the first $\sim$400~Myr despite the very large atmospheric losses.
This is due to the weak fractionation for atmospheres with very high loss rates meaning that hydrogen and oxygen atoms are lost at a 2:1 ratio.
After this phase, the mass loss decreases leading to a large amount of O$_2$ remaining in the atmosphere, and by 1~Gyr it is possible that $\sim$300~bar of O$_2$ is accumulated. 
The mass fractionation in the slow rotator case is much more efficient early on due to the higher fractionation factors, but due to the much weaker overall mass loss, approximately the same amount of O$_2$ can be accumulated. 
Interestingly, the medium rotator case leads to the most significant accumulation of O$_2$ since this case the star's activity spends much of the first Gyr at levels high enough to cause high escape rates while still low enough to allow fractionation factors below unity. 
The assumption that the star remains saturated for the first Gyr leads to no O$_2$ build-up in the atmosphere since the mass loss rate remains high for the entire time.
In all cases, the mass loss and the accumulation of O$_2$ can be limited if the reservoir of H$_2$O available to the atmosphere is exhausted.
This is most likely to happen for the fast rotator case and we might expect a much smaller accumulation of O$_2$ to take place; in this case, no O$_2$ accumulation might take place if the water in the atmosphere is entirely removed within the first 400~Myr and no significant amount of water is released into to the atmosphere afterwards.


\begin{figure*}
\centering
\includegraphics[width=0.45\textwidth]{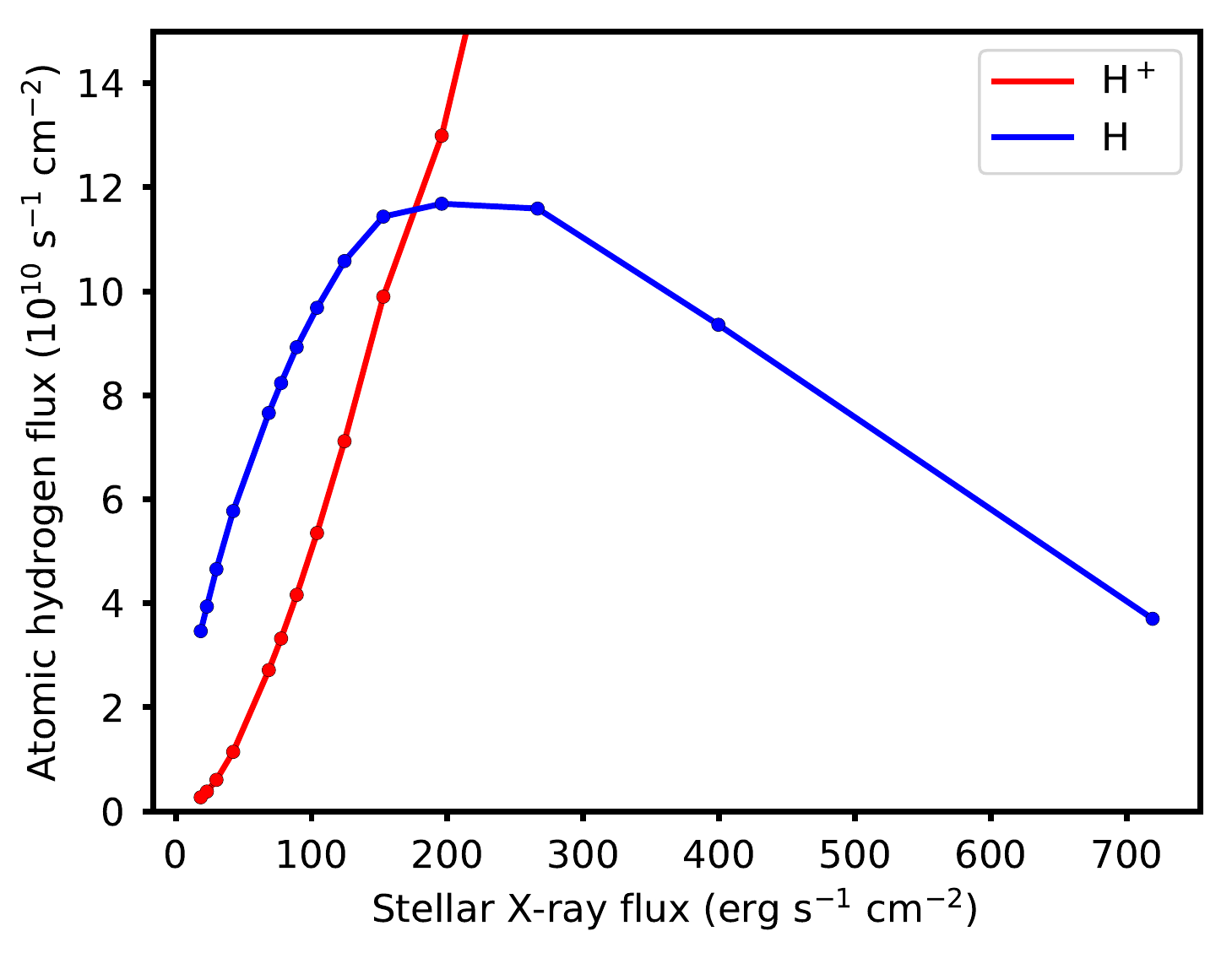}
\includegraphics[width=0.45\textwidth]{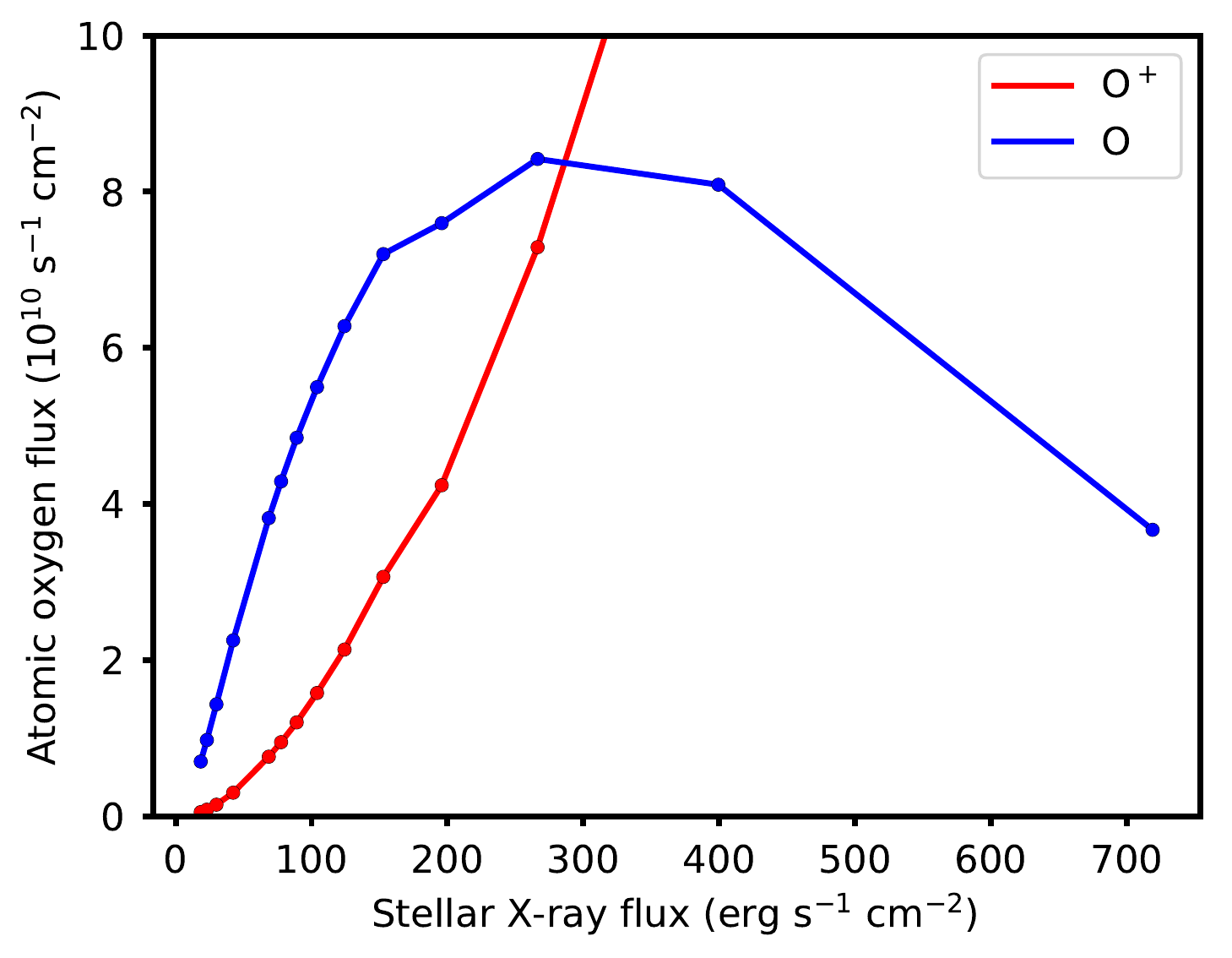}
\caption{
Outward fluxes for neutral and ionized hydrogen (\emph{left-panel}) and oxygen (\emph{right-panel}) as functions of stellar input X-ray flux.
}
\label{fig:HOlossGrid}
\end{figure*}


\section{Discussion} \label{sect:conclusions}

In this paper, I use a sophisticated physical model for the upper atmospheres of planets to study the hydrodynamic losses of water vapor atmospheres on Earth mass planets orbiting active stars with a range of activity levels.
In all cases studied, the heating of the atmosphere is strong enough to cause strong hydrodynamic escape leading to rapid water loss from the atmosphere.
Since the water molecules are dissociated by XUV photoreactions, the outflowing material consists of atomic hydrogen and oxygen, which in the most active cases and mostly ionized.
For the strongest outflows, I find that there is no separation of the particles by mass in the outflow, meaning that the losses will not lead to a build-up of oxygen or a change in the atmosphere's D/H ratio.

Connecting these results to evolutionary tracks for stellar XUV evolution, I study how much water vapor can be lost and how this might change the atmosphere's composition.
The results depend sensitively on the star's activity evolution which is determined by its mass and initial rotation rate of the star.
For solar mass stars, I find that a fast rotator can remove several tens of Earth oceans of water in the first billion years if a large enough reservoir of water is available, whereas a slow rotator can remove $\sim$1~Earth ocean.
The two cases however lead to very similar total amounts of oxygen accumulated, though if there is only a limited reservoir of atmospheric water vapor, then this build-up will be limited for the fast rotator case. 
In both cases, $\sim$300~bar of O$_2$ can be accumulated, whereas for the medium rotator case, a large amount of $\sim$ 450~bar is possible.
This O$_2$ will not necessary remain in the atmosphere, but can be absorbed into the planet's surface (\citealt{Wordsworth18}), lost hydrodynamically when it starts to become abundant enough in the upper atmosphere (\citealt{Tian15}), or lost later by other processes such as stellar wind pick-up (\citealt{Kulikov07}).
I also study the case of a star that remains at the saturation threshold, which is possible for fully convective M-dwarfs which remain highly active for very long time periods, and in some cases even several billion years (\citealt{West08}).
In this case, very large amounts of water can be lost from their atmospheres, but no build-up of O$_2$ takes place since the loss takes place rapidly enough that there is no significant mass fractionation in the outflow.
This means that the very large amount of O$_2$ accumulation that has been suggested for planets orbiting in the habitable zones of M-dwarfs (e.g. \citealt{LugerBarnes15}) could be unreaslistic.

It is important to consider the possible sources of error in my results.
While the model is the most sophisticated general purpose aeronomy model that has applied to problems of exoplanetary upper atmospheres, there are a very large number of improvements that are possible.
An obvious improvement would be the extension of the model to 3D geometries. 
While 1D models are able to accurately reproduce the upper atmospheres of solar system terrestrial planets, suggesting that considering more dimensions might by unnecessary, 3D models would be able to take into account the effects of different planetary rotation rates and planetary magnetic fields.
Given that in the most active cases that I consider in this paper, the gas becomes almost entirely ionized, strong interactions with the planet's intrinsic magnetic field (if present) might be possible, which has been shown to be able to drastically reduce mass loss rates for hot Jupiters (e.g. \citealt{Khodachenko15}).
Another planned improvement to the model is the implementation of a more sophisticated model for atmospheric cooling which could change the mass loss rates.
Finally, it is important to note that I have only considered the reaction of an atmosphere to stellar spectra of active solar mass stars.
While planets in the habitable zones of active lower mass stars, such as fully convective M stars, will receive similar amounts of total X-ray and EUV radiation, they will have different spectral shapes (\citealt{Fontenla16}).
The most important difference is that the photospheric spectrum, which for a G star dominates the XUV emission at wavelengths longer than $\sim$160~nm, is shifted to much longer wavelengths for M stars.
This could be important since this radiation is strongly absorbed by O$_2$ and O$_3$ molecules, which are created by the photodissociation of H$_2$O.

It is often stated in the literature that the photodisociation and heating of a water vapor outflow leads to a hydrodynamic outflow of H that drags O with it.
This is a misleading interpretation of what is taking place and implies that thermal pressure gradients only accelerate the hydrogen, while the oxygen atoms are only accelerated because they collide with the already accelerated hydrogen atoms.
In reality, the acceleration takes place where the atmosphere is still dense and collisional, and both the hydrogen and oxygen atoms are accelerated by the thermal pressure of the surrounding gas.
Atomic oxygen accelerates away from the planet not because it is `dragged' by hydrogen, but because it is accelerated by the same pressure gradients that cause the acceleration of hydrogen. 
It should be noted that at no point in the atmosphere is the gas dominated by atomic hydrogen: although H is the most numerous species above the lower thermosphere, O contributes most of the mass.

In all cases considered, the XUV heating of the gas, the acceleration of the hydrodynamic flow, the dissociation of H$_2$O, and the separation of the various species by mass by molecular diffusion happen all within a very small range of altitudes.
A major difference between hydrostatic and hydrodynamically outflowing atmospheres is that above this region in hydrodynamic atmospheres, the outward advection flow dominates over molecular diffusion.
This drastically limits how much molecular diffusion can separate particles by mass, leading to weak fractionation, and for the most active cases, there is no mass fractionation. 
The weak isotopic fractionation of hydrogen caused by extreme hydrodynamic escape is consistent with the results of \citet{Kasting83}. 
While they did not study such high solar activity levels, they found decreasing mass fractionation of hydrogen with increasing activity level (see their Fig.~13).

The strong outflows that I have calculated will lead to planets that are surrounded by a large cloud of particles (\citealt{Kislyakova13}; \citealt{Bourrier15}).
If this cloud of particles is transiting the star, the escape might lead to observable signatures that could be used to constrain the properties of the outflow and of the atmosphere.
The most obvious of these is transit observations in the host star's Ly-$\alpha$ line (\citealt{Kislyakova19}), and such signatures have already been seen in several systems (e.g. \citealt{Ehrenreich08}; \citealt{Kislyakova14b}; \citealt{Lavie17}).
In this case, the star can be transited by neutral hydrogen atoms flowing out of the planet's atmosphere and by energetic neutral atoms (ENAs) created by charge exchange between neutral atmospheric particles and stellar wind protons.
In both cases, a supply of neutral hydrogen from the planet's atmosphere is needed.
In Fig.~\ref{fig:HOlossGrid}, I show the outflow rates of neutral hydrogen and oxygen atoms.
For low input XUV fluxes, the neutral hydrogen and oxygen outflows increase rapidly with increasing XUV flux; however, this breaks down at very high stellar activities due to the increasing ionization of the gas.
For the most active cases, higher stellar activity leads to fewer neutral atoms flowing away from the planet, suggesting that the very high ionization fractions for the highest stellar activities might limit our ability to observe the outflow using Ly-$\alpha$ absorption.

Given how weak the isotopic fractionation of hydrogen by rapid hydrodynamic escape is, it is interesting to consider how much water must have been lost in order to produce the very large D/H ratio found on Venus (e.g. \citealt{Grinspoon93}; \citealt{Johnson19}). 
The total initial inventory of deuterium needed to produce the current D/H ratio in the atmosphere of Venus is given by \mbox{$N_\mathrm{D}^0 = N_\mathrm{D} \left( R/R_0 \right)^{1/(1-f_\mathrm{H,D})}$}, where $N_\mathrm{D}$ and $N_\mathrm{D}^0$ are the current and initial inventories of D respectively, and $R$ and $R_0$ are the current and initial D/H ratios respectively. 
Although the D/H ratio of the source of water in Venus' atmosphere is not known, it was likely much smaller than the current value, meaning that the term \mbox{$R/R_0$} should be very large. 
Assuming an $R_0$ similar to that of the modern Earth gives \mbox{$R/R_0 \sim 100$}.
In the lowest activity case calculated here, the value for $f_\mathrm{H,D}$ is 0.86, which gives \mbox{$1/(1-f_\mathrm{H,D})=7.2$} and means that an unrealistically large initial inventory of water must have been lost from Venus' atmosphere if extreme hydrodynamic escape was the primary mechanism for D/H enrichment.
Instead, it is likely that the large D/H ratio is a consequence of water loss that took place after the Sun's activity had declined significantly, which could have been either after the first few hundred million years if the Sun was a slow rotator or the first billion years if it was a fast rotator. 
The dominant loss process was likely either a much weaker hydrodynamic escape or non-thermal escape processes.
This does not rule out the possibility that rapid hydrodynamic water loss took place early in Venus' evolution and if so, it could have resulted in massive losses of water without any build-up of O$_2$ in the atmosphere.

\section{Acknowledgments} 

This study was carried out with the support by the FWF NFN project S11601-N16 ``Pathways to Habitability: From Disk to Active Stars, Planets and Life'' and the related subproject S11604-N16. 

\bibliographystyle{apj}
\bibliography{mybib}

\end{document}